\documentclass{iopjournal}
\usepackage{times} 
\usepackage{amssymb,amsmath}
\usepackage{csquotes}
\usepackage{algorithm, algpseudocode}
\usepackage{longtable, array, placeins}

\usepackage{xcolor}
\usepackage[most]{tcolorbox}
\usepackage{listings}

\definecolor{white}{rgb}{1,1,1}
\definecolor{mygreen}{rgb}{0,0.4,0}
\definecolor{light_gray}{rgb}{0.97,0.97,0.97}
\definecolor{mykey}{rgb}{0.117,0.403,0.713}

\tcbuselibrary{listings}
\newlength\inwd
\newcounter{ipythcntr}
\renewcommand{\theipythcntr}{\texttt{[\arabic{ipythcntr}]}}

\newtcblisting{pyin}[1][]{%
  sharp corners,
  enlarge left by=\inwd,
  width=\linewidth-\inwd,
  enhanced,
  boxrule=0pt,
  colback=light_gray,
  listing only,
  top=0pt,
  bottom=0pt,
  overlay={
    \node[
      anchor=north east,
      text width=\inwd,
      font=\footnotesize\ttfamily\color{mykey},
      inner ysep=2mm,
      inner xsep=10pt,
      outer sep=0pt
      ] 
      at (frame.north west)
      {\refstepcounter{ipythcntr}\label{#1}In~\theipythcntr:};
  }
  listing engine=listing,
  listing options={
    aboveskip=1pt,
    belowskip=1pt,
    basicstyle=\linespread{0.8}\footnotesize\ttfamily,
    language=Python,
    keywordstyle=\color{mykey},
    showstringspaces=false,
    stringstyle=\color{mygreen}
  },
}
\newtcblisting{pyprint}{
  sharp corners,
  enlarge left by=\inwd,
  width=\linewidth-\inwd,
  enhanced,
  boxrule=0pt,
  colback=white,
  listing only,
  top=0pt,
  bottom=0pt,
  overlay={
    \node[
      anchor=north east,
      text width=\inwd,
      font=\footnotesize\ttfamily\color{mykey},
      inner ysep=2mm,
      inner xsep=10pt,
      outer sep=0pt
      ] 
      at (frame.north west)
      {};
  }
  listing engine=listing,
  listing options={
      aboveskip=1pt,
      belowskip=1pt,
      basicstyle=\linespread{0.8}\footnotesize\ttfamily,
      language=Python,
      keywordstyle=\color{mykey},
      showstringspaces=false,
      stringstyle=\color{mygreen}
    },
}
\newtcblisting{pyout}[1][\theipythcntr]{
  sharp corners,
  enlarge left by=\inwd,
  width=\linewidth-\inwd,
  enhanced,
  boxrule=0pt,
  colback=white,
  listing only,
  top=0pt,
  bottom=0pt,
  overlay={
    \node[
      anchor=north east,
      text width=\inwd,
      font=\footnotesize\ttfamily\color{mykey},
      inner ysep=2mm,
      inner xsep=10pt,
      outer sep=0pt
      ] 
      at (frame.north west)
      {\setcounter{ipythcntr}{\value{ipythcntr}}Out#1:};
  }
  listing engine=listing,
  listing options={
      aboveskip=1pt,
      belowskip=1pt,
      basicstyle=\footnotesize\ttfamily,
      language=Python,
      keywordstyle=\color{mykey},
      showstringspaces=false,
      stringstyle=\color{mygreen}
    },
}

\begin{document}

\articletype{Paper}

\title{Hyperparameter Optimization in the Estimation of PDE and Delay-PDE models from data}

\author{Oliver Mai$^{1, *}$\orcid{0009-0007-5864-7674}, Tim W. Kroll$^1$, Uwe Thiele$^{1,2,3}$\orcid{0000-0001-7989-9271} and Oliver Kamps$^{3}$\orcid{0000-0003-0986-0878}}

\affil{$^1$Institute of Theoretical Physics, University of M\"unster, Wilhelm-Klemm-Str.\ 9, 48149 M\"unster, Germany}

\affil{$^2$Center for Data Science and Complexity (CDSC), University of M\"unster, Corrensstr.\ 2, 48149 M\"unster, Germany}

\affil{$^3$Center for Multiscale Theory and Computation (CMTC), University of M\"unster, Corrensstr.\ 40, 48149 M\"unster, Germany}

\affil{$^*$Author to whom any correspondence should be addressed.}

\email{oliver.mai@uni-muenster.de}


\begin{abstract}
We propose an improved method for estimating partial differential equations and delay partial differential equations from data, using Bayesian optimization and the Bayesian information criterion to automatically find suitable hyperparameters for the method itself and also for the equations (such as a time-delay). We show that combining time integration into an established model estimation method increases robustness and yields predictive models. Allowing hyperparameters to be optimized as part of the model estimation results in a wider modeling scope. We demonstrate the method's performance for a number of synthetic benchmark problems of different complexity, representing different classes of physical behavior. This includes the Allen-Cahn and Cahn-Hilliard models, as well as different reaction-diffusion systems without and with time-delay. \\
The published version of this preprint can be found under this \href{https://iopscience.iop.org/article/10.1088/1367-2630/ae2e33}{link} or the DOI: 10.1088/1367-2630/ae2e33.
\end{abstract}

\section{Introduction} \label{sec:intro}

The recent advancements in data collection and storage, as well as the rising availability of computational resources, have facilitated data-driven learning methods and impacted modern science in many ways. As large-scale experiments and computer simulations continue to produce complex data sets, typical methods for discovering governing physical laws and translating them into mathematical models, such as first-principle derivations, take more time and effort to complete or may not account for all the available data. Thus over recent years computer-aided discovery of dynamical equations has attracted more and more attention.

During their inception many symbolic data-driven modeling methods focused on ordinary differential equations (ODEs). As a results there are numerous such methods available, e.g., the multiple shooting method~\cite{Bock1984}, sensitivity analysis~\cite{Guay1995} or generalized Gauss-Newton methods~\cite{Houska2012}. Then, reference~\cite{Brunton2016} proposed to reframe an inverse problem in terms of a least-square problem with a suitable set of candidate model functions. This spawned various alterations and in recent developments the combination of various different methods. This includes combining de-noising and data preparation into the learning method~\cite{Egan2024}, or incorporating methods that allow for domain-specific user input to cope with sparsely sampled data~\cite{Omejc2024}.

However, for many (especially spatially extended) systems a description using ODEs is insufficient and partial differential equations (PDEs) are needed. Although, in principle, methods originally designed for ODEs can readily be adapted for use with PDEs, often new challenges arise regarding scalability and performance. Attempts to specifically identify PDEs from data started in the late 1990s~\cite{Br1999, Ljung1998, Voss1999, Voss1998} and have seen various improvements and modifications over the years. 
Typically, just as for ODEs, a model structure is either proposed by prior knowledge about the system or expressed by a set of candidate functions, which are then narrowed down using, e.g., sparsity-promoting algorithms~\cite{Brunton2016, Rudy2017, VOSS2004, Wang2019}, sensitivity-analysis~\cite{Zhao2020, Naozuka2022} and information criteria \cite{Mangan2017, Dong2022}. The models are optimized to either minimize the residual of the PDE~\cite{Brunton2016, Rudy2017, Rudy2019} or its predictive error~\cite{Long2019, Fullana1997}.
While there have been variations and alternative methods, where the system is described via a weak-formulation~\cite{Messenger2021, Tang2023}, methods which allow for additional complexity such as time-delayed variables~\cite{Sandoz2023, ddefind2024} remain few and narrow in scope.

A parallel development is found in physics-informed neural networks (PINNs)~\cite{zubov2021neuralpde} and related approaches. There, knowledge about the underlying physical systems is incorporated to subsidize otherwise black-box approaches to predict system behavior or to recover specific system parameters, but not to generate a comprehensive mathematical description of the provided data.

For many of the model learning methods the sole benchmark remains the rediscovery of physical laws based on numerical simulations. The performance on these benchmarks has made significant progress for both ODEs and PDEs. While this facilitates the use of these methods for real world applications, practical application to study open problems still faces many challenges. Examples include microscopy data~\cite{Maddu2022}, gene expression~\cite{Sandoz2023} and monitoring of motors~\cite{Koch2021} or tokamak discharge~\cite{Wan2021}, but poor knowledge of involved parameters and overfitting onto noise still pose significant hurdles. The model quality is often only evaluated after the learning method has concluded~\cite{Dong2022}. This results in models that are numerically unstable during time simulations or fail to replicate initial data at all, without additional clean-up or corrections using prior knowledge.

In many data-driven model estimation methods, hyperparameters that govern model complexity, regularization, data preprocessing or optimization strategies have a vital influence on the performance of the learning algorithms~\cite{Zhao2018, Machlanski2023}. These may gauge overfitting or set learning rates or thresholds during optimization. Often these hyperparameters are not learned from data, but instead set beforehand and either adjusted via a trial-and-error approach or via computationally expensive systematic grid-based or random searches.

In reference~\cite{Kroll2025} the REBEL (Reconstruction Error Based Estimation of dynamical Laws) method, a novel method for identifying sparse ODE systems from data, is introduced. It combines the efficient sparse parameter estimation approach of reference~\cite{Brunton2016} with an error estimate based on a combination of the integrated least squares error and the Wasserstein metric. Integrated within a Bayesian optimization framework, this method efficiently determines optimal hyperparameters, leading to a better data approximation with fewer parameters.
Our goal is to leverage this approach for robust and efficient sparse PDE model estimation from data. The hyperparameter optimization strategy in reference~\cite{Kroll2025} also enables us to efficiently estimate delay PDEs, thereby significantly expanding the applicability of the method.

In section~\ref{sec:num_methods} we develop our numerical method. Then, in section~\ref{sssec:baseline} we show that it yields virtually identical results for the elementary example analyzed in reference~\cite{Rudy2017}. The same holds true in section~\ref{sssec:space_steps_and_noise}, where we compare resilience to noise and low spatial sample rate.
To further investigate the comparative performance, we vary the sampling frequency in section~\ref{sssec:var_sample} and show that our approach is more robust than the one in reference~\cite{Rudy2017} with respect to data that are under-sampled in time. Then, in section~\ref{ssec:pattern_formation} we consider phase-field models with and without mass-conservation and show that further restrictions are not needed to achieve mass-conservation and that our implementation allows more flexibility for model ansatz functions. 
Finally, section~\ref{ssec:hyperparameter} showcases how the method accommodates multiple hyperparameters, such as multiple method thresholds and time-delays, which enables us to treat models of higher complexity. section.~\ref{sec:outlook} summarizes our findings and proposes further fields of study. The Appendix houses more detailed information on creating a default set of ansatz functions in section~\ref{apx:generating_library}, a pseudo-code representation of the model estimation procedure in section~\ref{apx:pseudo_code}, an in-depth numerical tutorial in section~\ref{apx:che_tutorial} and unabbreviated numerical results for all examples shown in section~\ref{apx:results}.

\section{Numerical Methods}  
\label{sec:num_methods}
\subsection{Optimization framework}

The symbolic approach to estimating dynamic equations in the form of PDEs from data relies on narrowing down suitable functions for the right hand sides of the equations from a set of ansatz- or library-functions. In its simplest form this means taking time series data and then fitting the library-functions to the time derivative of the data. In practice, when using a least-squares fit, this often results in extensive and unwieldy expressions, where most of the functions only give small contributions. The aim is to single out only major contributions and to produce well-established and/or interpretable models.

We assume that there exists a functional expression $F$, which describes the true dynamics of the time evolution of a physical quantity $u(x,t)\in \mathbb{R}^{n}$ (where $n$ is the number of field variables) we wish to model:
\begin{equation}
    \frac{\partial u}{\partial t} =  F (u).
\end{equation}
$t$ is the time and $x\in \mathbb{R}^d, d\in \{1, 2, 3\}$ position in $d$-dimensional space. $F (u)$ may include nonlinearities and spatial derivatives of $u$ and in principle any function that contributes to the dynamics of $u$. We assume that each individual contribution in $F$ (however complex it may be) can be combined linearly, so $F$ could look as follows:
\begin{align}
    \frac{\partial u}{\partial t} &= a u + b u^2 + c \nabla u + d \sin(t) + \ldots,  \label{eqn:demo_lib}
\end{align}
where $a, b, c$ and $d$ are coefficients. Now, for an unknown system it may be unclear which exact functions appear on the right-hand side of equation~(\ref{eqn:demo_lib}), but many scientific disciplines hold broad catalogs of modeling approaches. As such we assemble a large number of ansatz- or library-functions that could capture the dynamics of an unknown system we wish to model (see  Appendix~\ref{apx:generating_library} for more details on library creation). We collect all of the library-functions in a vector $\Theta$
(with $k$ number of functions) and all of their coefficients in a matrix $\sigma \in \mathbb{R}^{n \times k}$, so that we can rewrite equation~(\ref{eqn:demo_lib}) as the linear matrix multiplication
\begin{equation}\label{eqn:inverse_problem}
    \frac{\partial u}{\partial t} =  \sigma \cdot \Theta (u).
\end{equation}
Of course, if the true dynamics falls outside of the space spanned by our library functions, then they can only be approximated.

\begin{figure}
    \centering
    \includegraphics[width=0.9\textwidth]{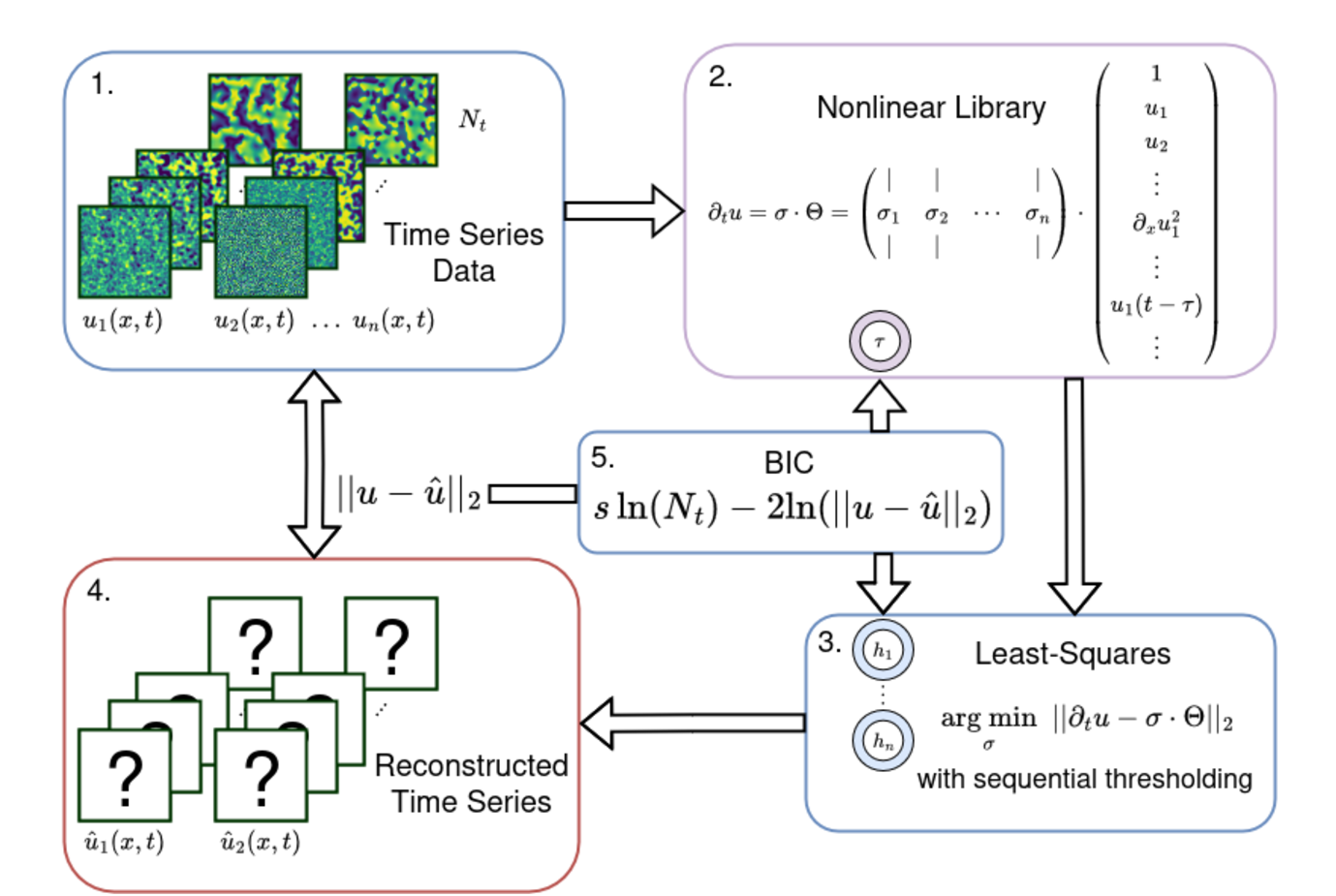}
    \caption{Scheme of the model estimation procedure. The steps are:
    1. load time series data;
   	2. compute suitable library functions;
   	3. use least-squares-based sparse regression method to estimate model parameters;
   	4. integrate estimated model in time;
   	5. compute BIC and repeat steps 2 (if library terms have changed), 3 and 4.}
    \label{fig:estimation_routine}
\end{figure}

The goal is now to find a sparse set of values for $\sigma$ that best describes the time evolution of $u$. To that end, we employ a regression method. See figure~\ref{fig:estimation_routine} for an overview over the entire optimization framework.
So we search for an approximate set of coefficients, that has a residual error with respect to the true system:
\begin{equation}\label{eqn:residual}
    R(\sigma) = || \partial_t u - \sigma \cdot \Theta (u)||_2.
\end{equation}
Here,
\begin{equation}
||z||_2 = \left(\sum_{r=1}^{N_t} \sum_{l=1}^{N^2} |{z_r}_l|^2\right)^{\frac{1}{2}}
\end{equation}
 is the $\ell^2$-norm for $N_t$ number of samples in time and $N^2$ sample points in space. Now we want to find the set $\hat{\sigma}$, that minimizes $R$. In other words, we wish to solve the optimization problem
\begin{equation}\label{eqn:minimize}
   \hat{\sigma} =  \underset{\sigma}{\text{arg min}}~~ R(\sigma) = \underset{\sigma}{\text{arg min}}~~ || \partial_t u - \sigma \cdot \Theta (u)||_2.    
\end{equation}
Since $\sigma \cdot \Theta$ is a linear matrix multiplication, this problem can be numerically solved by any least-squares routine.

This leaves two problems. First, as previously mentioned, a least-square routine by itself typically yields non-sparse results. Second, the residual in equation~(\ref{eqn:residual}) is formulated with respect to the time derivative of the original data and as shown in reference~\cite{Kroll2025} the minimum of such a residual is not identical to the one using the deviation from the integrated estimate to the original time series:
\begin{equation}\label{eqn:likelihood}
	L(\hat{\sigma}) = \Big|\Big| u - \int_0^T \hat{\sigma} \cdot \Theta(u) \text{d}t \Big|\Big|_2 = || u - \hat{u}||_2.
\end{equation}

The use of numerical time integration (from $t=0$ to the final time $t=T$) means, that a reconstructed time series $\hat{u}$, now only implicitly depends on our coefficients $\hat{\sigma}$.

To overcome both of these problems we employ the same strategy as reference~\cite{Kroll2025}, i.e., we make use of the Bayesian Information Criterion (BIC), which takes the maximized likelihood $\tilde{L}$ of our estimated model (which is equivalent to minimizing equation~(\ref{eqn:likelihood})) and penalizes it with the number of model parameters:
\begin{equation}\label{eqn:BIC}
    \text{BIC} = s \ln(N_t) - 2 \ln(\tilde{L}).
\end{equation}
Here $s$ is the number of non-zero coefficients in $\hat{\sigma}$.
When $\hat{\sigma}$ is close (or equal) to the true system parameters, $L(\hat{\sigma})$ should also be minimal, thus $L(\hat{\sigma}) = \tilde{L}$. This results in the scheme shown in figure~\ref{fig:estimation_routine}:
\begin{enumerate}
 \item Load time series data of $u(x,t)$ that shall be modeled.
 \item Construct the library $\Theta$ of nonlinear ansatz functions which may contain hyperparameters such as a time delay $\tau$.
 \item Perform a least-squares fit to find estimated system parameters $\hat{\sigma}$ and sequential thresholding with hyperparameters $h_1, \ldots, h_n$, where contributions of $\hat{\sigma}_i$ to the evolution of variable $u_i$ below the threshold $h_i$ are subsequently discarded until sparsity no longer increases.
 \item Perform time integration using the estimated parameters $\hat{\sigma}$ and library $\Theta$.
 \item Use the deviation from the original time series and the BIC to optimize the hyperparameters and repeat steps~2 (if any library term has changed due to a hyperparameter), 3 and 4.
\end{enumerate}
In other words, we begin by using a sparsity promoting least-squares routine to find an initial $\hat{\sigma}$, which minimizes the residual~(\ref{eqn:residual}) with respect to the time derivative, then compute $L$ using $\hat{\sigma}$ (and any additional model parameters) by integrating our model in time and finally use another numerical optimization procedure to iterate previous steps to minimize the BIC. This yields optimized method hyperparameters, such as any thresholds $h$ in this sparsity promoting variant of the least-squares routine, and refined model parameters. Additionally, we may add any other hyperparameter to be optimized, such as ones that change our ansatz functions (e.g., a time delay $\tau$ or a phase shift $\varphi$).  For more details on the entire optimization routine see again figure~\ref{fig:estimation_routine} and section~\ref{ssec:estimation_procedure}. 

\subsection{Estimation Procedure}\label{ssec:estimation_procedure}
Even for a large library $\Theta(u)$, problem~(\ref{eqn:inverse_problem}) remains a linear matrix equation. Thus we initially estimate parameters  $\hat{\sigma}$ using the least-squares routine from the popular computing package \emph{Numpy}~\cite{harris2020}, minimizing the difference to our time derivative:
\begin{equation*}
    \hat{\sigma} = \underset{\sigma}{\text{arg min}}~~ || \partial_t u - \sigma \cdot \Theta (u)||_2.    
\end{equation*}
The time derivative of the original time series data $\partial_t u$ is computed using finite differences. Spatial derivatives during library evaluation can be computed via FFT or finite differences (supplied by the package \emph{findiff}\cite{findiff}). Our routine uses the latter per default. The accuracy of these derivatives can greatly impact the estimation performance, albeit with diminishing returns: Falling below a certain accuracy will result in a worse estimation performance depending on the problem at hand. Higher accuracy, however, will not necessarily increase the estimation performance. The accuracy of the finite differences method can be freely adjusted. The default is second order accuracy. 

Sparsity of $\hat{\sigma}$ is then promoted using sequential thresholding least-squares~(STLS, \cite{Brunton2016}), i.e., parameters with a magnitude below a threshold value $h$ are set to zero and excluded, then the least-square fit is repeated until either sparsity no longer increases or a maximum number of iterations is reached.
The threshold $h$ constitutes a hyperparameter that greatly influences the resulting model. In reference~\cite{Rudy2017} the STLS algorithm is used with one identical threshold value for all parameters. In contrast, here, we allow for a threshold $h_i$ per system variable $u_i$. Additionally, one may want to group thresholds by the influence of certain groups of library terms. For example when adding time-delayed library terms, we add a threshold $h_{\tau}$, for those terms.

To find these threshold hyperparameters we use the optimization package \emph{Hyperopt}~\cite{Bergstra2015} and its tree-structured Parzen estimator (TPE)~\cite{Bergstra2011}, which is commonly used for hyperparameter tuning in machine learning. The TPE is a Bayesian optimization method that models a given objective function probabilistically. It compares the ratio of two density estimators to focus on promising regions of the search space. We set the objective function of the TPE to be the BIC~(\ref{eqn:BIC}). This requires the computation of equation~(\ref{eqn:likelihood}) and therefore $\hat{u}$, which we obtain by integrating the model in time. This time integration makes use of the initial value problem solver of \emph{SciPy}~\cite{SciPy2020} and its explicit Runge-Kutta method of order 8~\cite{Hairer1993}, although any method from SciPy may be selected (e.g., an implicit method). Such time integration can be a difficult task, however, since in the process of finding a feasible PDE description, in some cases, candidate PDEs do not guarantee numerical stability. Hence we utilize a trapezoidal integration scheme (also the one implemented in SciPy~\cite{SciPy2020}) as a fallback method when integration fails, or to reduce computation time. While different numerical time integration methods can be used, as in the case of the spatial derivatives. While more stable or higher-order integrators may enable successful estimation where simpler methods fail, they introduce increased computation time and may not necessarily improve estimation performance otherwise.

Additionally we may feed any other hyperparameter that defines our ansatz library into the TPE. One example is the later described time-delay $\tau$ for certain terms. In these cases values of the functions in $\Theta$ change, so they are recomputed during the optimization if needed.
Again, for an overview of the estimation routine see figure~\ref{fig:estimation_routine} or the pseudo-code representation in algorithm~\ref{alg:TSME} found in  Appendix~\ref{apx:pseudo_code}.

\section{Results}
\label{sec:results}
All results in this section (if not otherwise stated) have been achieved using the Python package \emph{TSME} (\lq time series model estimation\rq, software publication pending, preliminary release in reference~\cite{TSME}). All relevant data and Python code for the examples given here can be found in the \href{https://github.com/CeNoS-CoSy/tsme_examples}{GitHub repository} in reference~\cite{TSMEgithub}.

This section investigates several synthetic inverse problems to illustrate the performance of the developed method. First, we consider a symmetric two-field reaction-diffusion model equivalent to the complex Ginzburg-Landau equation, which was used as a benchmark in reference~\cite{Rudy2017}. In particular, the goodness of fit is investigated when varying the sample frequency in space and time or with added noise. As the software package \emph{PySINDy}~\cite{deSilva2020} that originally implemented the method proposed in reference~\cite{Brunton2016} was later extended for use with PDEs via the method introduced in reference~\cite{Rudy2017}, that implementation will be used as a reference. The second case is the Cahn-Hilliard equation, a model with mass conservation for which corresponding state-of-the-art software packages are to our knowledge not yet applicable, due to their restrictive ansatz library. The final three examples focus on the application of multiple hyperparameters. The FitzHugh-Nagumo reaction-diffusion model and the chaotic regime of the complex Ginzburg-Landau equation are employed to illustrate how incorporating semantic information into the choice of multiple thresholds may lead to improved results. Finally we show for the Fisher-KPP equation with added time delay how the method can incorporate additional hyperparameters, aside from thresholds. While all examples are based on synthetic data, they serve as proof-of-concept and show that for cases where underlying equations are unknown the inclusion of additional hyperparameters strongly improves model estimation abilities. 

\subsection{Dependence of Performance on Data Quality}
\subsubsection{Baseline Performance}\label{sssec:baseline}
As in reference~\cite{Rudy2017} we generate synthetic data for a reaction-diffusion model with third-order nonlinearities corresponding to a complex Ginzburg-Landau (cGL) equation, namely:
\begin{align}
    \begin{split}
        \partial_t u &= D_u \nabla^2 u + (1-A)u + Av \label{eqn:rd_1} \\
        \partial_t v &= D_v \nabla^2 v + (1-A)v - Au \\
        A &= u^2 + v^2.
    \end{split}
\end{align}
We use a two-dimensional spatial domain with coordinates $x$ and $y$ each ranging from $-10$ to $10$ (thus with domain size $L_x \times L_y = 20\times 20$) and periodic boundary conditions. The initial condition corresponds to a spiral centered at the origin given by
\begin{align}
	\begin{split}
		\label{eqn:rd_ic}
		u_0(x, y) &= \tanh(r) \, \cos\!\big(\theta - r\big), \\
		v_0(x, y) &= \tanh(r) \, \sin\!\big(\theta - r\big),\\
\text{where}\quad		r = \sqrt{x^2 + y^2}, &\quad \theta = \arg(x + i y).
	\end{split}
\end{align}

In terms of a linear combination of polynomial functions in the fields and their derivatives the system reads
\begin{align}
    \begin{split}
        \partial_t u &= D_u \partial_x^2 u + D_u \partial_y^2u + u -u^3 - uv^2 + u^2v + v^3 \\
        \partial_t v &= D_v \partial_x^2 v + D_v \partial_y^2v + v -v^3 - u^2v - uv^2 - u^3.  
    \end{split} 
\end{align}

\begin{figure}
    \centering
    \includegraphics[width=0.9\textwidth]{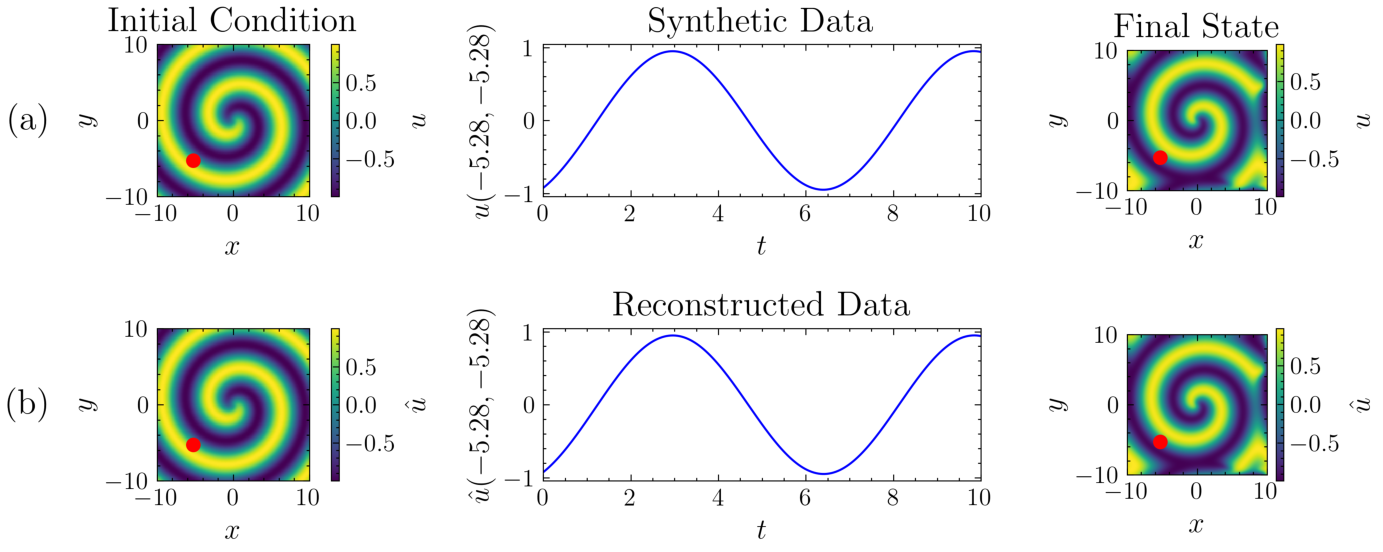}
    \caption{Time simulation of equations~(\ref{eqn:rd_1}) with $D_u = D_v = 0.1$  (first row) and of the data-based estimated equations~(\ref{eqn:rd_id_1}) (second row). In both cases the spatial grid is $128 \times 128$, the domain size is $L_x=L_y = 20$ and the time step is $0.05$. Images on the left show the identical initial conditions, images on the right show the final snapshot. The graphs in the center give the time trace at the location marked in red within the snapshots. As the spiral rotates the traces oscillate. Very good agreement between the original synthetic data and their reconstruction with the estimated model is found.}
    \label{fig:rd_and_estimate_selection}
\end{figure}

Synthetic data are generated using a time simulation of equations~(\ref{eqn:rd_1}) with initial condition~(\ref{eqn:rd_ic}). The $u$-component is shown in the first row of figure~\ref{fig:rd_and_estimate_selection}.
Here and in the following, results shown for PySINDy have been produced using the corresponding Python package~\cite{deSilva2020}. For the sake of comparability, here, we use our method with only one threshold $h$ and with the identical library as employed in reference~\cite{deSilva2020}, as described in Appendix~\ref{apx:ssec_library_sindy}.

With a maximum order $m=3$ of power products and a maximum order $p=2$ for spatial derivatives both, our method implemented in TSME using the BIC (equation~(\ref{eqn:BIC})) as well as the one implemented in reference~\cite{deSilva2020} (PySINDy) identify the system as:
\begin{align}
    \begin{split}
        \partial_t \hat{u} &= 0.10 \partial_x^2 \hat{u} + 0.10 \partial_y^2\hat{u} + 0.96\hat{u} -0.96\hat{u}^3 - 0.96\hat{u}\hat{v}^2 + 1.00\hat{u}^2\hat{v} + 1.00\hat{v}^3 \label{eqn:rd_id_1} \\
    \partial_t \hat{v} &= 0.10 \partial_x^2 \hat{v} + 0.10 \partial_y^2\hat{v} + 0.96\hat{v} -0.96\hat{v}^3 - 0.96\hat{u}^2\hat{v} - 1.00\hat{u}\hat{v}^2 - 1.00\hat{u}^3 
    \end{split}
\end{align}
(here and in the following results are rounded to the second decimal place, see  Appendix~\ref{apx:results} table~\ref{tab:apx_rd} for complete results).
The time simulation for this reconstructed model is given in the second row of figure~\ref{fig:rd_and_estimate_selection}. Both methods correctly identify the underlying equations out of 110 candidate functions with a single-digit error percentage ($\approx 4.09\,\%$) for the prefactors. While for PySINDy the threshold from the original reporting in reference~\cite{Rudy2017} is used ($h=0.08$), our optimization procedure found a threshold $h\approx 0.0812$. Note, that reference~\cite{Rudy2017} seems to have determined $h$ heuristically, while the here developed optimization method automatically chooses $h$ to minimize the BIC~(\ref{eqn:BIC}). 

\subsubsection{Influence of Spatial Discretization and Noise}\label{sssec:space_steps_and_noise}
\begin{figure}
    \centering
    \includegraphics[width=0.8\textwidth]{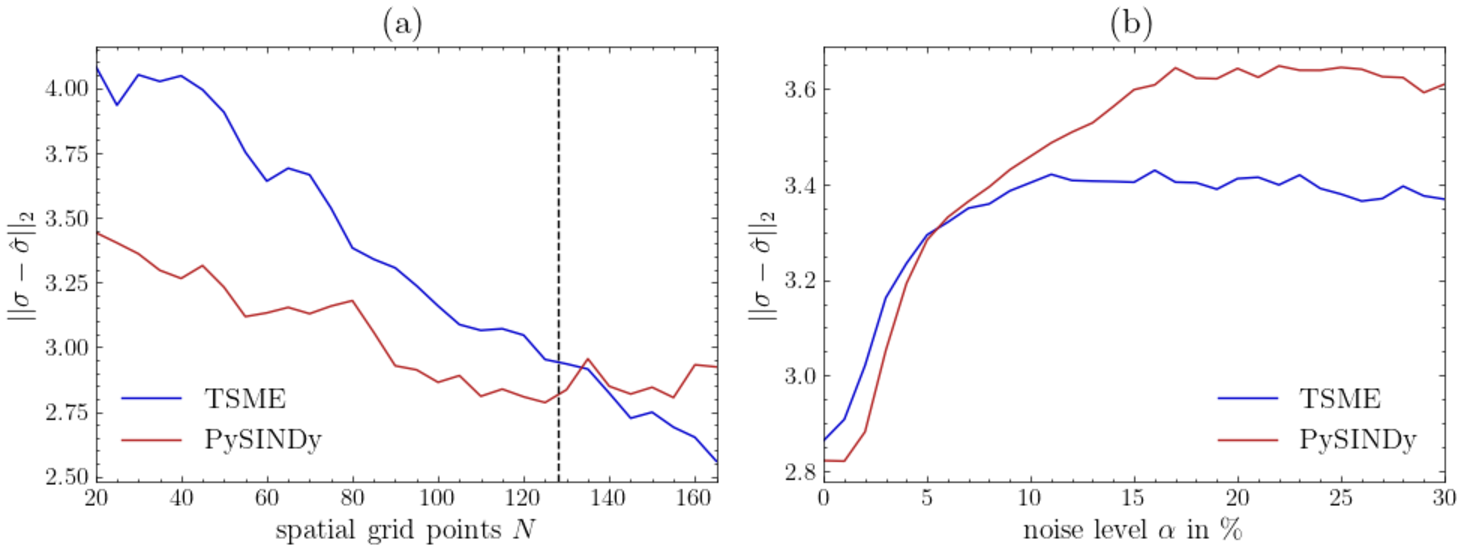}
    \caption{
    Average $\ell^2$-error of the model parameters $||\sigma - \hat{\sigma}||_2$ over 10 data series for (a) varying spatial sampling rate $N$ and without additive noise and (b) for additive noise of various strength $\alpha$ at fixed $N=128$. In panel (a) $N$ increases from $20$ to $165$ in increments of $5$ and in panel (b) $\alpha$ increases from $0\,\%$ to $30\,\%$ in increments of $1\,\%$. The vertical dashed line in panel (a) indicates the origin of panel (b) at $N=128$ and $\alpha=0$.}
    \label{fig:error_vs_space_noise}
\end{figure}

Here, we first briefly compare the performance of our method implemented in TSME and and the method of reference~\cite{Rudy2017} implemented in PySINDy~\cite{deSilva2020} with respect to the spatial sampling rate $N$ for the $N \times N$ spatial grid of data points. We again perform time simulations for equations~(\ref{eqn:rd_1}) with identical model parameters, time interval and boundary conditions, but for varying $N$ and with random initial conditions, where for each variable all $N^2$ values are sampled from a uniform distribution over the interval $[-0.25, 0.25)$. Thereby, PySINDy optimizes only with respect to the residual $R=||\partial_t u - \hat{\sigma}\cdot \theta||_2$  ($h=0.08$) and TSME instead incorporates the BIC (in both cases we use the library from section~\ref{sssec:baseline} with $m=3, p=2$) averaged over 10 model estimations for each $N$. Figure~\ref{fig:error_vs_space_noise}~(a) shows the error of the estimated model parameters. We see that both methods are similarly susceptible to a low spatial sampling rate and have a performance that improves with increasing $N$. While TSME is not as good as PySINDy for small $N$ it shows a stronger improvement with increasing $N$ and is the better method at larger $N$. 

Second, we compare the performance of TSME and PySINDy with respect to their resilience to noisy data, in particular, we fix $N=128$ and add noise to all input data using
\begin{equation}
	\tilde{u} = u\left(1 + \alpha\, \eta\right),
\end{equation}
where $\eta$ is a random real number sampled from a normal distribution with mean value $0$ and standard deviation $1$ and $\alpha$ is a scaling factor that gauges noise strength. Otherwise, our methodology remains unchanged, we take the same model parameters, time interval, boundary and initial conditions as before and perform 10 model estimations for each of various $\alpha$. The results are shown in figure~\ref{fig:error_vs_space_noise}~(b). We see that the performance of both methods suffers after only a few percent of noise added. While the overall error of our method is lower for stronger noise, both TSME and PySINDy fail to reliably recover the correct model after approximately 5\,\%.
This implies that a de-noising of input data is advisable when dealing with noisy data. Alternatives are briefly discussed in the conclusion.

\subsubsection{Varying Sample Frequency in Time}\label{sssec:var_sample}
\begin{figure}
    \centering
    \includegraphics[width=0.8\textwidth]{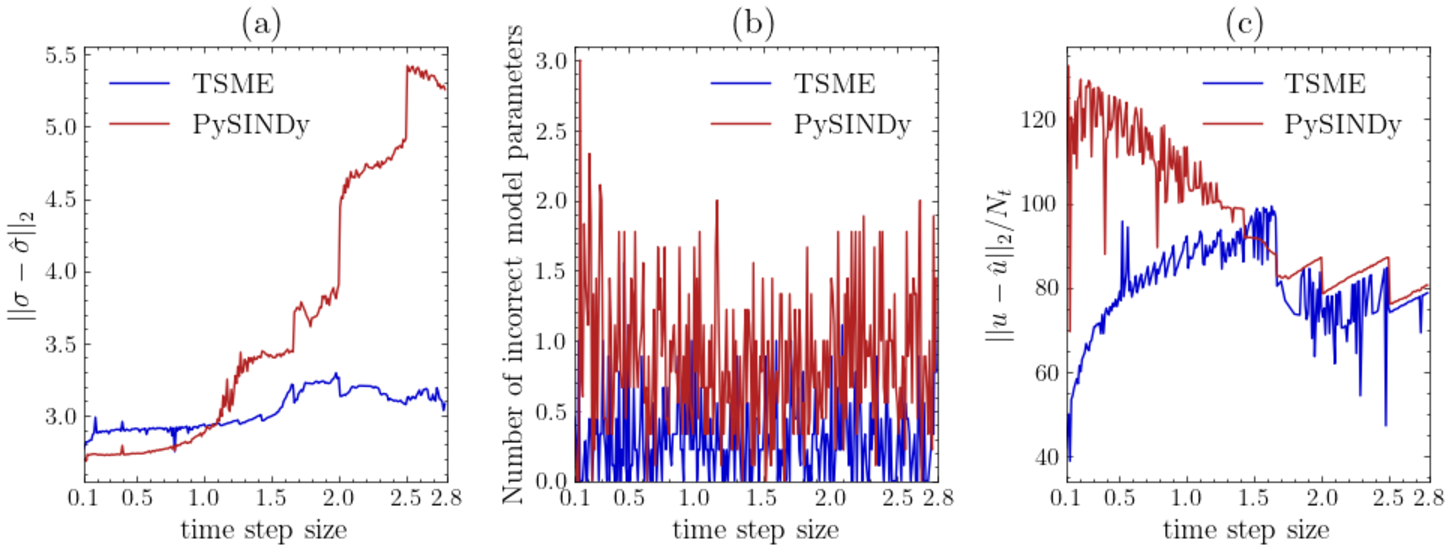}
    \caption{
    Average estimation error over 10 data series with random initial conditions for varying time step size, ranging from $0.1$ to $2.8$ in increments of $0.01$. Panel (a) shows the average $\ell^2$-error of the model parameters $||\sigma - \hat{\sigma}||_2$, panel (b) gives the average number of falsely identified model parameters, which may be a positive real number due to averaging, panel (c) shows the average root mean squared error (RMSE) of the reconstructed time series using the estimated model $||u - \hat{u}||_2 / N_t$. The proposed BIC error shows a slightly increased deviation in its model parameters for lower time step sizes, but greatly reduced error for larger time step sizes. Sparsity of the BIC error consistently outperforms the error w.r.t. only the time derivative}
    \label{fig:error_vs_time_step_size}
\end{figure}

Here, we investigate the performance of TSME and PySINDy for different time step sizes of the data. Time simulations are performed for equations~(\ref{eqn:rd_1}) with the previous model parameters and random initial conditions without additive noise at fixed $N=128$. The resulting figure~\ref{fig:error_vs_time_step_size} shows (a) the average deviation of the estimated model parameters from their true value, (b) the average number of incorrectly identified model parameters, and (c) the average deviation of a reconstructed time series using the estimated model from the original time series. 

As we can see in figure~\ref{fig:error_vs_time_step_size} (a) for sufficiently small time steps both methods yield parameters very close to their true values. Thereby, PySINDy shows a smaller deviation from the parameter values for smaller time steps, but also less sparsity across all time step sizes (i.e., more incorrectly identified model parameters in figure~\ref{fig:error_vs_time_step_size} (b)) and thus a greater error in the reconstructed time series (figure~\ref{fig:error_vs_time_step_size} (c)). So while the proposed optimization method deviates more from true model parameter values for smaller time step sizes, it identifies correct model terms more reliably and thus yields better results for time integration.It should be noted, that PySINDy contains various extensions to improve the performance when sub-sampling data or to accommodate poorly sampled data, which are not being used here. By simply incorporating time integration into these methods, one should be able to increase the robustness further.

Between the two methods the average number of incorrectly identified terms varies greatly, but seems to be independent of the time step size between samples. While the variance could be reduced by increasing the number of time series per sample, it remains unclear as to why sparsity does not diminish for increasing time step size as one would expect. Similarly, the errors between the estimated time series and the original one do not mirror the trend seen in the errors of the model parameters.

\subsection{Phase Field Models for Pattern Formation}
\label{ssec:pattern_formation}
Phase field models, such as the Allen-Cahn equation and the Cahn-Hilliard equation, provide a description of interface evolution and phase separation in various physical systems. The key distinction between these two models is their \enquote{mass-conservation character}: The Allen-Cahn equation describes the dynamics of a non-conserved order parameter, where the local field can change dynamically without restriction, e.g., in reaction to external fields. In contrast, the Cahn-Hilliard equation locally enforces mass conservation as it has the form of a continuity equation. This ensures that phase separation does not alter the total quantity of each material. The non-conserved case lends itself readily to regression methods for model estimation, whereas the conserved case often requires constrained optimization approaches. In this section we show that the here proposed optimization method can directly be applied in both cases.   
\subsubsection{Allen-Cahn equation}\label{ssec:allen_cahn}
The Allen-Cahn equation has been widely used to model phase transitions across different scientific fields. In material science, it describes the phase separation in binary alloys and order-disorder transitions~\cite{Cahn1994}. As such it has become a basic model to study crystal growth and interfacial dynamics~\cite{Hussain2019, Stegemerten2020}. In biology it forms a basis for models simulating cancer cell migration, providing insights into tumor development and metastasis~\cite{Bandeira2024}.

We consider the following Allen-Cahn equation, that represents a gradient dynamics on a double-well potential:
\begin{equation}\label{eqn:ACE}
	\partial_t u = \nabla^2 u + u - u^3.
\end{equation}
Here, $u(x, t)$ is a non-conserved order parameter, typically representing a phase field of a bistable system. We rewrite it in terms of our default library as
\begin{equation}
	\partial_t u = \partial_{x}^2 u + \partial_{y}^2 u + u - u^3.
\end{equation}
A time simulation of equation~(\ref{eqn:ACE}) for periodic boundary conditions and uniform random initial conditions (over the interval $[-0.25, 0.25)$) can be found in the first row of figure~\ref{fig:ace_selection}.
Now we apply the BIC optimization method with the default library (see Appendix~\ref{apx:default_library}) with $m=3$ and $p=4$ (i.e., we allow for spatial derivatives up to 4th order and up to cubic nonlinearities encompassing in total 46 terms). We obtain
\begin{equation}\label{eqn:ACE_id}
	\partial_t \hat{u} = 0.99\partial_{x}^2 \hat{u} + 0.99\partial_{y}^2 \hat{u} + 0.99\hat{u} - 0.99\hat{u}^3
\end{equation}
(for complete results see Appendix~\ref{apx:results} table~\ref{tab:apx_ace}). A time simulation of the estimated model is found in the second row of figure~\ref{fig:ace_selection}. 

\begin{figure}
    \centering
    \includegraphics[width=0.9\textwidth]{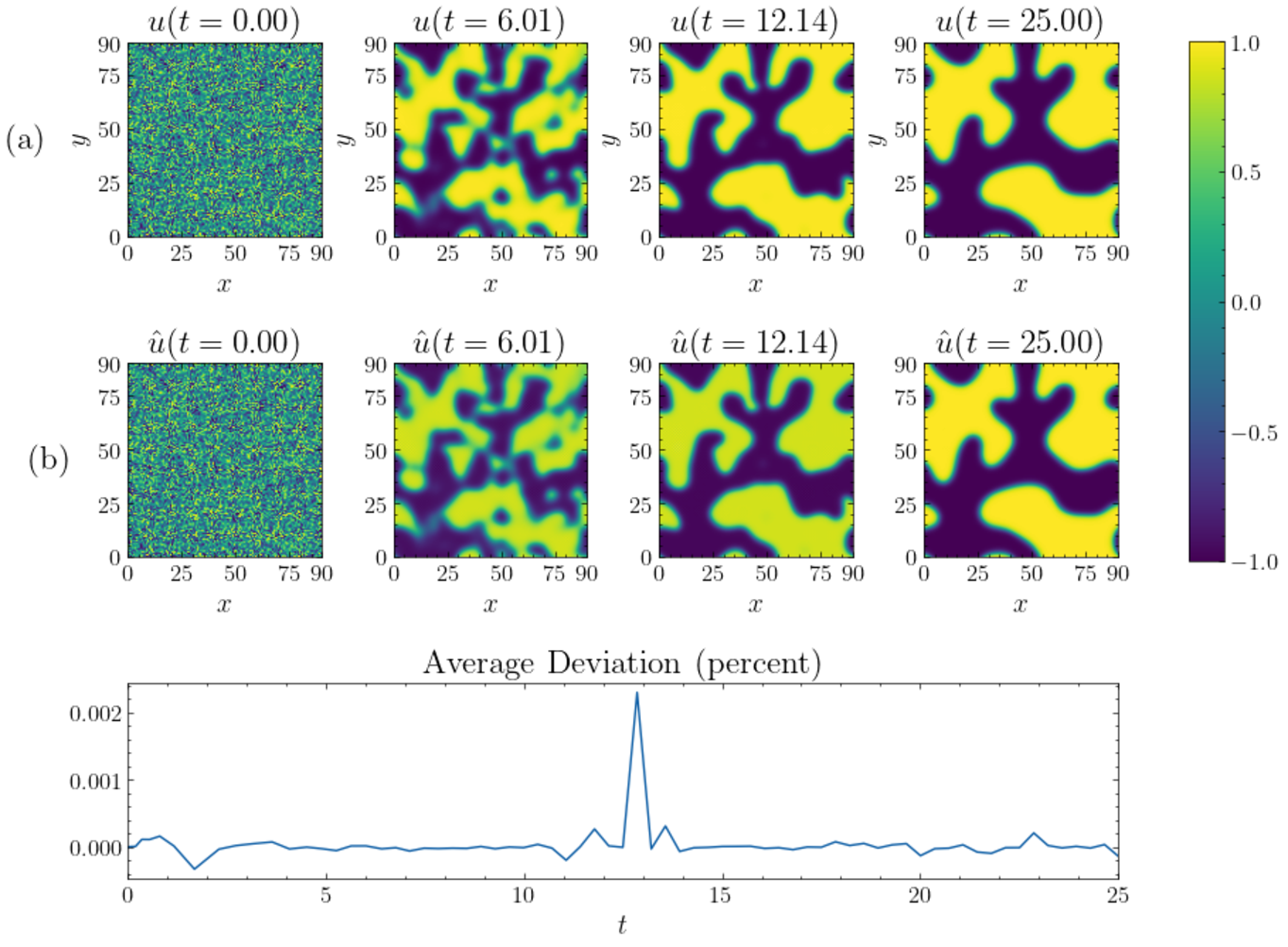}
    \caption{Time simulation of the original Allen-Cahn equation~(\ref{eqn:ACE}) (top row) and the estimated equation~(\ref{eqn:ACE_id}) (middle row) for a time step of $0.125$ (spatial domain $128 \times 128$ and $L_x=L_y=90$). The bottom row show the average deviation of the reconstructed time series in percent. The estimated model shows very good agreement.}
    \label{fig:ace_selection}
\end{figure}
Consistent with section~\ref{sssec:baseline} we find very good agreement between original and estimated model with errors below $2\,\%$. 

\subsubsection{Cahn-Hilliard equation}\label{ssec:cahn_hilliard}

\begin{figure}
    \centering
    \includegraphics[width=0.9\textwidth]{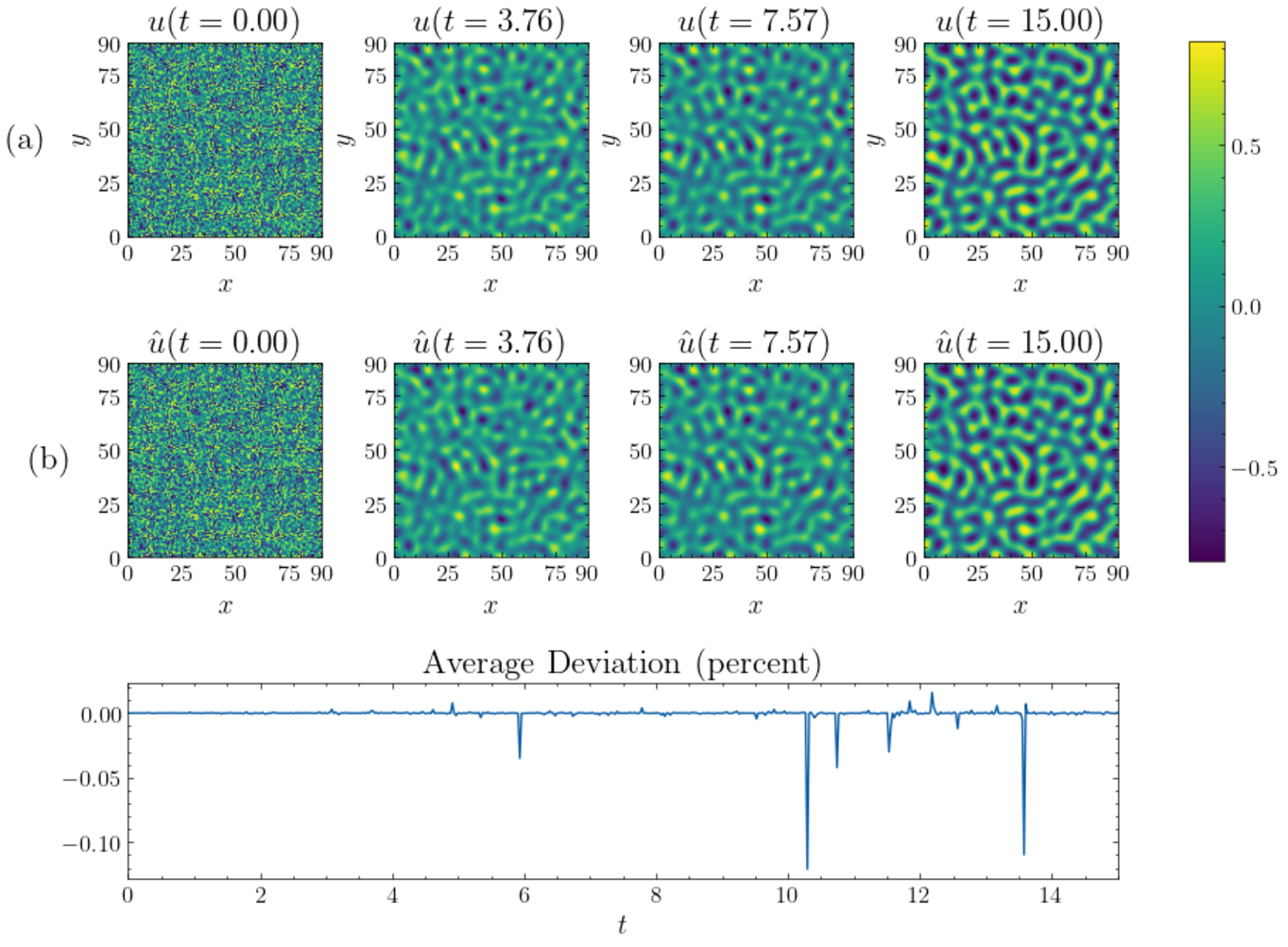}
    \caption{Time simulation of the original Cahn-Hilliard equation~(\ref{eqn:CHE}) (top row) and the estimated equation~(\ref{eqn:CHE_id}) (middle row) for a time step of $0.15$ (spatial domain $128 \times 128$ and $L_x=L_y=90$). The bottom row show the average deviation of the reconstructed time series in percent.}
    \label{fig:che_selection}
\end{figure}
Next, we consider spinodal decomposition in a binary mixture as described by the Cahn-Hilliard (CH) equation. Generally, the CH equation is used to study structure formation in phase-separating alloys or other multi-component systems, often in chemical or biological contexts~\cite{Kim2016}. Therefore, one may expect to find the CH equation or related models when investigating inverse problems involving demixing behavior. 

Here, we use the original equation
\begin{equation}\label{eqn:CHE}
    \partial_t u = \nabla^2 \left( u^3 - u - \nabla^2 u \right),
\end{equation}
where $u$ denotes a difference in concentration between two species. Here as before we consider periodic boundary conditions and uniform random initial conditions (over the interval $[-0.25, 0.25)$).

As noted in reference~\cite{Brunk_2023}, several points related to ill-posedness deserve attention when estimating PDE models of CH-type, i.e., gradient-dynamics systems characterized by an energy functional and a mobility function. Any method that identifies the governing PDE effectively recovers the chemical potential, which can, in principle, be integrated to obtain the corresponding energy functional. For such a mass-conserving dynamics the corresponding integration constant is arbitrary and of no importance. Likewise, a constant external chemical potential (equivalent to a linear term in the energy) merely acts as a Lagrange multiplier enforcing the mean value of the field, which, however, is fixed by the initial condition and can be read directly from the data. Furthermore, there is an ambiguity between energy functional and a mobility function that can individually only be estimated up to a relative factor. However, also this is only of limited physical consequence, as it corresponds to a choice of time and energy scales. If necessary, this can be resolved by analyzing the final steady state to extract the energy directly. Note, however, that the candidate library of terms must be expanded if the mobility is expected to depend on the field as, e.g., for thin-film equations \cite{CrMa2009rmp}.

Restated in terms of our default library again with $m=3$ and $p=4$ equation~(\ref{eqn:CHE}) reads:
\begin{equation}
    \partial_t u = -\partial_x^2 u - \partial_y^2 u + \partial_x^2 u^3 + \partial_y^2 u^3 - \partial_x^4 u -2\partial_x^2\partial_y^2 u  - \partial_y^4 u.
\end{equation}
 Note that nonlinearities with second-order spatial derivatives cannot be included using the implementation provided by PySINDy in reference~\cite{deSilva2020} as it lacks the ability to input user--defined library terms and terms like $\partial_x^2 u^3$ are not of the form used in the example of section~\ref{sssec:baseline}, i.e., they cannot be rewritten in terms that only include spatial derivatives of linear order of $u$. The implementation presented here allows the user to manipulate the default library. Furthermore, the default library already includes the nonlinear terms with second-order spatial derivatives. Utilizing this implementation we obtain as result
\begin{equation}\label{eqn:CHE_id}
    \partial_t \hat{u} = -0.99\,\partial_x^2 \hat{u} - 0.99\,\partial_y^2 \hat{u} + 0.98\,\partial_x^2 \hat{u}^3 + 0.99\,\partial_y^2 \hat{u}^3 - 1.02\,\partial_x^4 \hat{u} -1.92\,\partial_x^2\partial_y^2 \hat{u}  - 1.02\,\partial_y^4 \hat{u}, 
\end{equation}
(for a complete list see  Appendix~\ref{apx:results} table~\ref{tab:apx_che}). Again the method correctly identifies the original equation correctly from 45 candidate terms within single digit ($\lesssim 4\,\%$) error percentage, despite the presence of a conservation law and without further constraints on the library. For a Python tutorial on how to obtain these results please see  Appendix~\ref{apx:che_tutorial}.

\subsection{Excitable, Oscillatory, and Wave Propagation Systems}\label{ssec:hyperparameter}
Up to this point we only utilized a single hyperparemeter, i.e., one threshold for all variables in the STLS method. To illustrate the flexibility of our optimization approach we next study three systems, which inherently benefit from including more than one hyperparameter in the optimization procedure.
\subsubsection{FitzHugh-Nagumo equation}\label{ssec:fhn}

\begin{figure}[h!]
    \centering
    \includegraphics[width=0.9\textwidth]{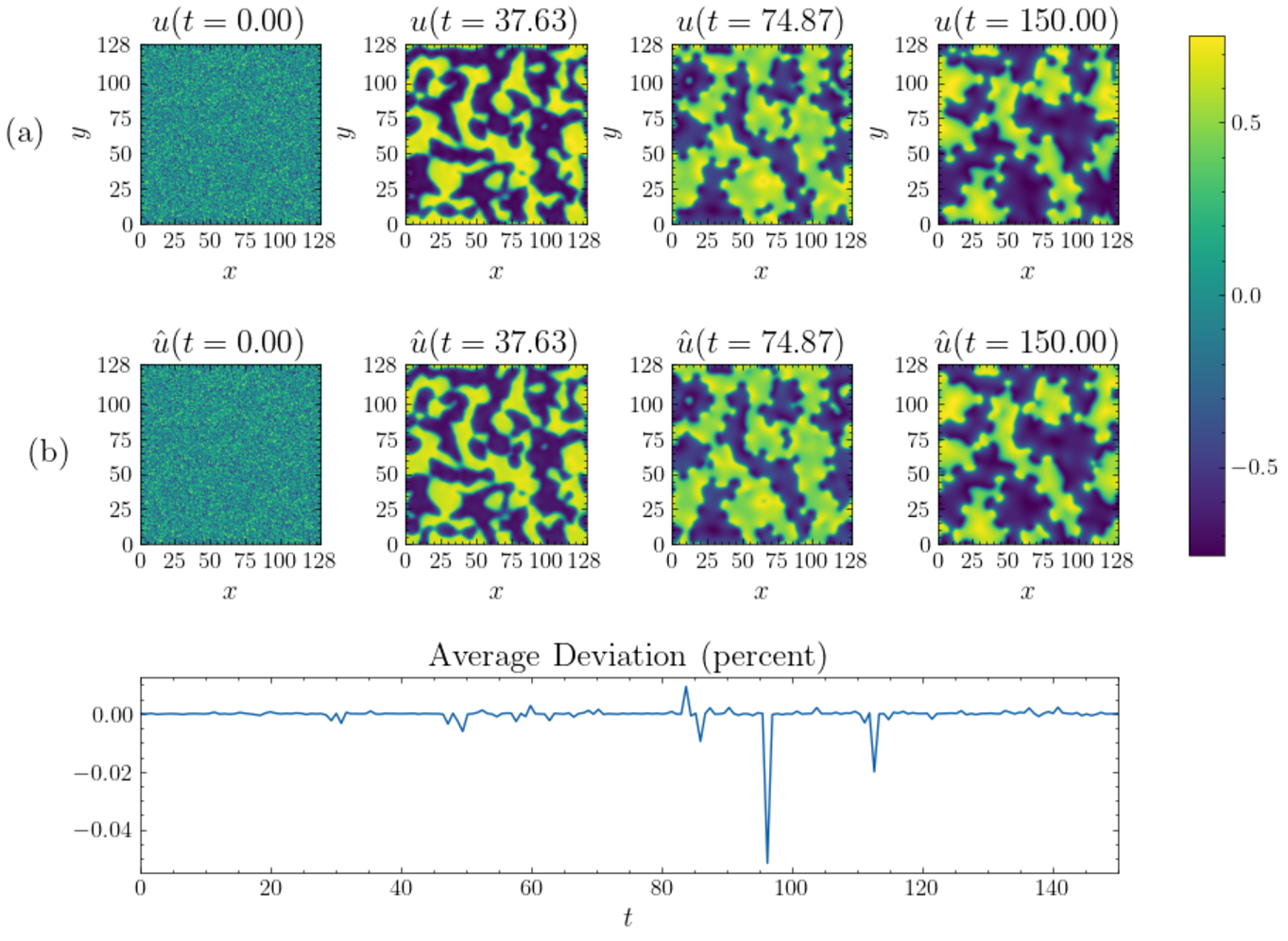}
    \caption{Snapshots from a time simulation of the original FHN equations~(\ref{eqn:FHN_u}) with $D_u =0.5$, $D_v = 0.0\overline{6}$ (top row) and the estimated equations~(\ref{eqn:FHN_id_u}) (middle row) at a time step of $0.75$ (spatial domain $256 \times 256$ and $L_x=L_y=128$). The bottom row show the average deviation of the reconstructed time series in percent.}
    \label{fig:fhn_estimate_selection}
\end{figure}

The FitzHugh-Nagumo (FHN) model~\cite{FitzHugh1961, Nagumo1962} is a two--species reaction-diffusion model central in investigations of the patterns of neuronal firing, synchronization of neural networks and the emergence of complex behavior in excitable media~\cite{Zheng2015, CebrinLacasa2024}. It is derived as a simplification of a model for the transmission of electrical impulses along a nerve fibre, as introduced by Hodgkin and Huxley~\cite{Hodgkin1952}. 
Due to its rich bifurcation behavior, the accurate estimation of the model parameters is vital to understanding how the states and patterns evolve in time. As such the FHN model is frequently considered in the context of parameter identification~\cite{Dong2015, Ahmed2021}.
Here we study the FHN model in the form
\begin{align} 
    \begin{split}
    \partial_t u &= D_u \nabla^2 u + \lambda u - u^3 - \omega v \\
    \tau\partial_t v &= D_v \nabla^2 v + u - v,
    \end{split}
\end{align}
where $u$ describes a membrane potential, $v$ a recovery variable, $\lambda$ a threshold for excitability, $\omega$ gauges inhibition strength, while $D_{u}$ and $D_{v}$ are respective diffusion constants. Again we use periodic boundary conditions and uniform random initical conditions (over the interval $[-0.25, 0.25)$).
The top row of figure~\ref{fig:fhn_estimate_selection} shows snapshots from a sample time series. 

We rewrite the model in terms of the adequate candidate functions that are a subset of our library and insert the specific parameters used in figure~\ref{fig:fhn_estimate_selection}:
\begin{align}
    \begin{split}
    \partial_t u &=  0.5u - 0.5v - 1.0u^3 + 0.5\partial_x^2u+ 0.5\partial_y^2 u \label{eqn:FHN_u} \\
    \partial_t v &= 0.0\overline{6} u - 0.0\overline{6} v + 0.0\overline{3}\partial_x^2 v + 0.0\overline{3}\partial_x^2 v, 
    \end{split}
\end{align}
Note that the order of magnitude of the parameters differs between the two equations. To account for this we enable individual thresholds $h_u$ and $h_v$ for each equation in the sequential thresholding procedure and optimize for both.
With our default library as described in Appendix~\ref{apx:default_library} with parameters $m=3$ and $p=2$, we identify the equations:
\begin{align}
    \begin{split}
    \partial_t \hat{u} &=  0.49 \hat{u} - 0.49\hat{v} - 0.97\hat{u}^3 + 0.24 \partial_x^2\hat{u} + 0.24 \partial_y^2\hat{u} \label{eqn:FHN_id_u} \\
    \partial_t \hat{v} &= 0.07 \hat{u} - 0.07 \hat{v} + 0.02 \partial_x^2 \hat{v} + 0.02 \partial_y^2 \hat{v},
    \end{split}
\end{align}
out of a library of 54 candidate terms for each equation (for full results see Appendix~\ref{apx:results} table~\ref{tab:apx_fhn}). We observe that all terms have been correctly identified, albeit with a diffusion coefficient for $\hat{u}$ that seems to be diminished by half. The obtained thresholds are $h_u=0.07$ and $h_v=0.01$. When using only one threshold, we find two cases. Either, the threshold is small enough, such that the second equation contains non-zero entries. However, in turn this yields non-zero contributions for virtually all available library functions in the first equation. Or, the threshold is large enough to lead to a sparse representation of the first equation, but in turn eliminates all library functions from the second equation. These results indicate that one needs several thresholds, when one expects groups of model parameters to differ in their orders of magnitude.  

\subsubsection{Exploring chaotic regimes}
\label{ssec:rd_chaotic}
Next, we revisit the cGL equation~(\ref{eqn:rd_1}) from section~\ref{sssec:baseline}, this time for a parameter regime where the system exhibits chaotic behavior like also considered in~\cite{Shraiman1992}. First, we show that our method maintains its performance even in a chaotic regime.

Second, we highlight the benefit of incorporating additional hyperparameters, namely thresholds for sub-groups of library terms as described in section~\ref{ssec:estimation_procedure}. Here we choose a separate threshold for all the diffusion terms.

To allow for different chaotic regimes we slightly modify the reaction--diffusion system, by additionally coupling the fields by cross diffusion terms, i.e., 
\begin{align}
    \begin{split}
    \partial_t u &= D_1 \nabla^2 u - D_2 \nabla^2 v + (1-A)u + \beta Av \label{eqn:rdc_1} \\
    \partial_t v &= D_1 \nabla^2 v + D_2 \nabla^2 u + (1-A)v - \beta Au \\
    A &= u^2 + v^2.
    \end{split}
\end{align}
Boundary conditions are periodic and initial conditions are uniform random (over the interval $[-0.25, 0.25)$).
With the particular choice of diffusion constants, the system is equivalent to the complex Ginzburg-Landau equation~\cite{Shraiman1992} that features different types of chaos. The added parameter $\beta$ serves as a way to control the coupling strength, as well as the phase rotation of a complex amplitude $w=u + i v$.

\begin{figure}
    \centering
    \includegraphics[width=0.95\textwidth]{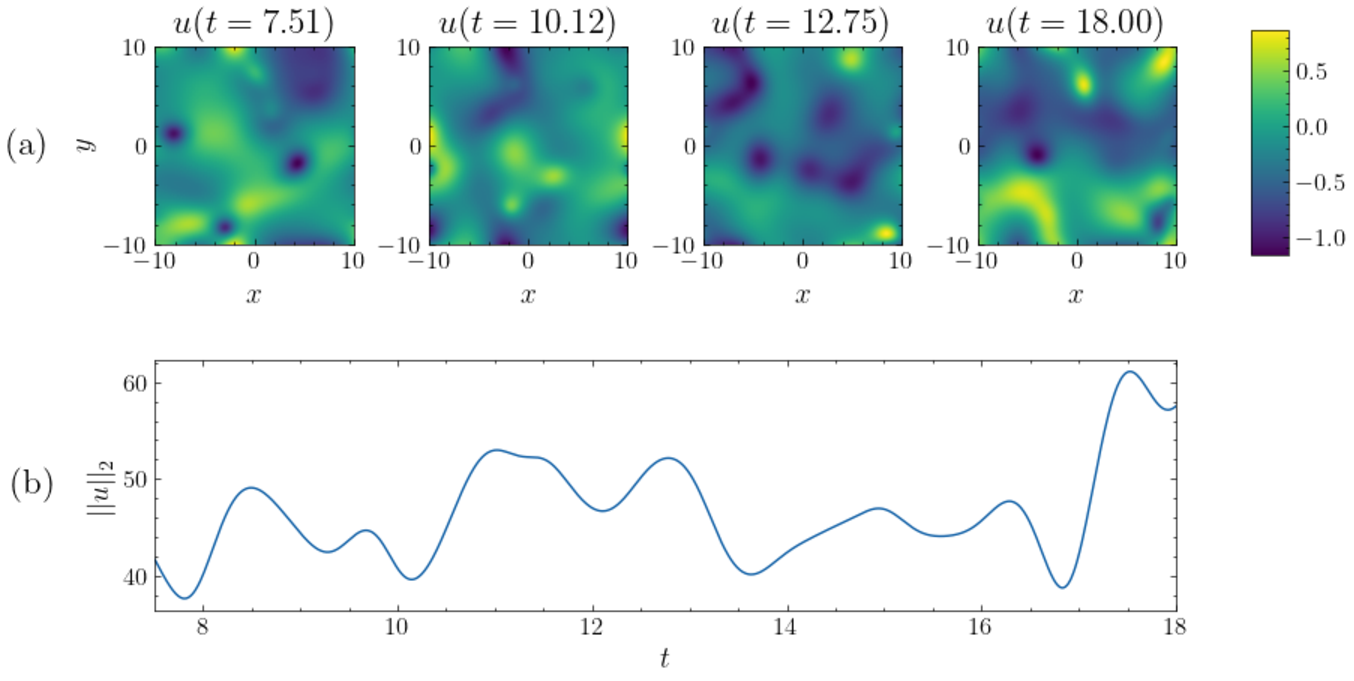}
    \caption{Time simulation in the defect turbulence regime of equations (\ref{eqn:rdc_1}) at $D_1=1.0$, $D_2=2.0$ and $\beta=-4$ with adaptive time step size between $0.004326$ and $0.017094$ (spatial domain $128 \times 128$ and $L_x=L_y=20$). The first row shows exemplary snapshots of $u$ and the second row shows the $\ell^2$-norm of $u$ over time. The initial condition for both $u$ and $v$ is white noise between values $-0.5$ and $0.5$.}
    \label{fig:rd_chaotic_1}
\end{figure}

\begin{figure}
    \centering
    \includegraphics[width=0.95\textwidth]{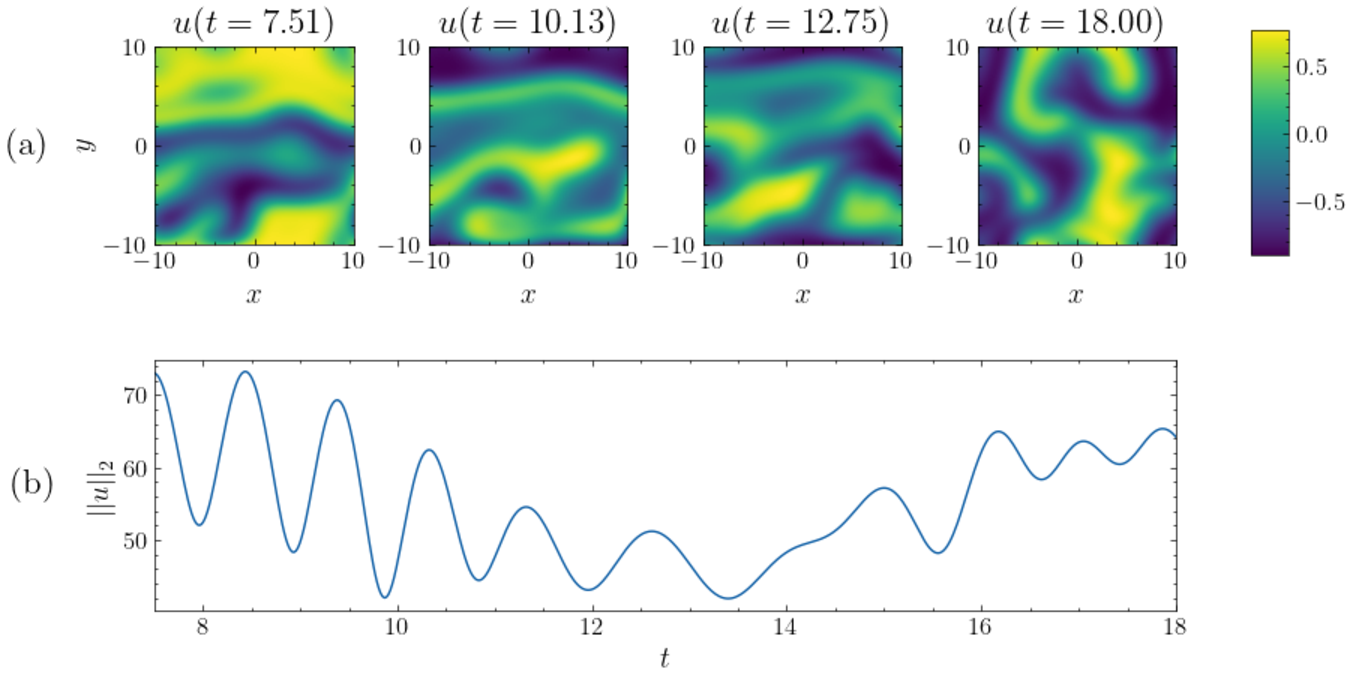}
    \caption{Time simulation in the intermittency regime of equations (\ref{eqn:rdc_1}) at $D_1=1.0$, $D_2=0.0$ and $\beta=-4$ with adaptive time steps between $0.0073265$ and $0.020553$ (spatial domain $128 \times 128$ and $L_x=L_y=20$). The first row shows exemplary snapshots of $u$ and the second row shows the $l^2$-norm of $u$ over time. The initial condition for both $u$ and $v$ is white noise between values $-0.5$ and $0.5$.}
    \label{fig:rd_chaotic_2}
\end{figure}

Figure~\ref{fig:rd_chaotic_1} shows a time simulation of equations (\ref{eqn:rdc_1}) for $D_1=1, D_2=2, \beta=-4$, which corresponds to a chaotic defect turbulence regime. Eliminating the cross coupling terms at otherwise identical parameters places the system in an intermittency regime with more coherent structures, as illustrated in figure~\ref{fig:rd_chaotic_2}. 

Applying the estimation method with a single thresholding hyperparameter in the defect turbulence regime very closely recovers the system parameters:
\begin{align}
    \begin{split}\notag
        \partial_t u &= 1.00 \partial_x^2 u + 1.00 \partial_y^2u
    - 2.00 \partial_x^2 v - 2.00 \partial_y^2 v
    + 1.00u -1.00u^3 - 1.00uv^2 - 4.00u^2v -4.00v^3 \\
    \partial_t v &= 1.00 \partial_x^2 v + 1.00 \partial_y^2v
    + 2.00 \partial_x^2 u + 2.00 \partial_y^2 u
    + 1.00v - 1.00v^3 - 1.00u^2v + 4.00uv^2 + 4.00u^3
    \end{split}
\end{align}
(for full results see  Appendix~\ref{apx:results} table~\ref{tab:apx_rd_chaotic_1}).
In contrast, when applying the method in this way in the intermittency regime, it either fails to eliminate higher-order diffusion terms (see terms $44$, $47$, $50$ and $53$ in table~\ref{tab:apx_rd_chaotic_2a} of  Appendix~\ref{apx:results}) or incorrectly eliminates all diffusion terms. The problem is resolved when making use of the methods ability to include additional hyperparameters. We incorporate two additional thresholds, one for each of the two pairs of second order diffusion terms (corresponding to $\nabla^2u$ and $\nabla^2v$). 
This dramatically improves results in case of the intermittency regime (see  Appendix~\ref{apx:results} table~\ref{tab:apx_rd_chaotic_2b} for full results):
\begin{align}
    \begin{split}
    \partial_t u &=  0.97\partial_x^2 u +  0.90\partial_y^2u
    + 0.96u + 0.96u^3 - 0.96uv^2 + 3.99u^2v - 3.99v^3 \label{eqn:rdc_id2_1} \\
    \partial_t v &= 0.97\partial_x^2 v + 0.90\partial_y^2
    + 0.96v -0.96v^3 - 0.96u^2v + 3.99uv^2 + 3.99u^3. 
    \end{split}
\end{align}
This behavior may be due to rapid changes of large coherent structures in the input time series, akin to the time series rapidly oscillating. This, in turn, effectively decreases spatial resolution of diffusive boundaries and produces dynamics that are less dependent on diffusion, leading to difficulties when estimating diffusion coefficients. 

\subsubsection{Time delay: Fisher-KPP equation with time delay}\label{ssec:fisherkpp}

The efficient handling of additional hyperparameters opens a straight forward way to handle delay equations. To our knowledge this has not yet been addressed in the framework of sparse system identification. As example we consider the Fischer-Kolmogorov-Petrovsky-Piskunov (Fisher-KPP) equation~\cite{Fischer1937, kolmogorov1937}. It is a one-species reaction-diffusion system used, e.g., to study propagation fronts of biological populations. Here, we include a time delay $\tau$ in the reaction term as inspired by references~\cite{Ducrot2014, Zhang2025}. The equation then reads
\begin{equation}\label{eqn:fisher_kpp}
    \partial_t u(t) = D \partial_{x}^2 u(t) + ru(t)\left( 1 - \frac{u(t-\tau)}{K} \right),
\end{equation}
where $D$ is a diffusion coefficient, $r$ an intrinsic growth rate and $K$ a carrying capacity. The time delay $\tau$ represents a lag between the current population growth and past population pressure. This accounts for example for the time it takes for members of a given population to reach reproductive maturity or other delayed feedback introduced by external time delayed resources.

\begin{figure}
    \centering
    \includegraphics[width=\linewidth]{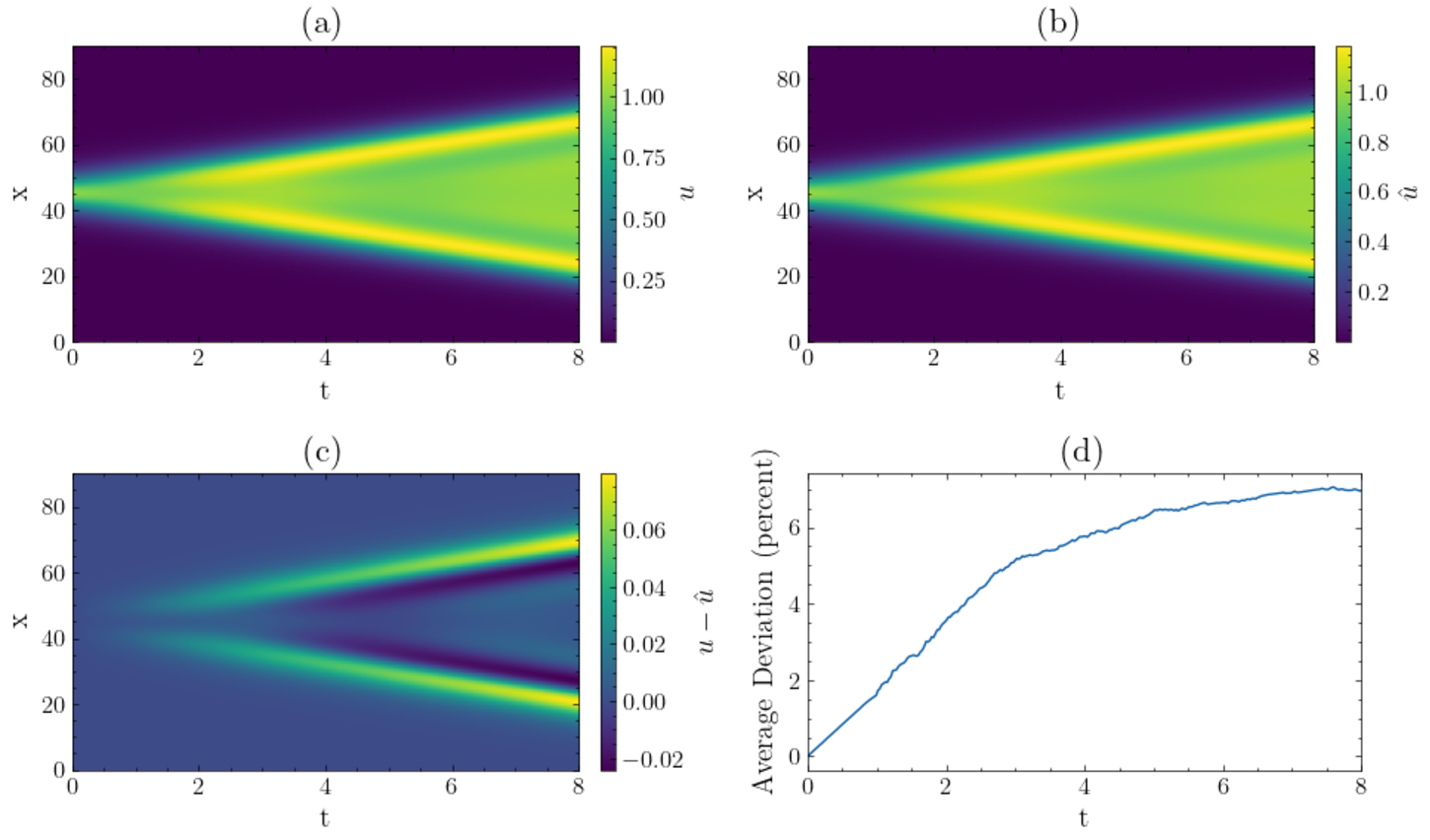}
    \caption{Time simulations of the FisherKPP equation~(\ref{eqn:fisher_kpp}), with all parameters set to $1.0$, time step size $0.0266$ (spatial discretization $128$, $L_x = 90$) (top left) and estimated model~(\ref{eqn:fisher_kpp_est}) (top right), as well as the absolute difference between the two (bottom left) and the average deviation of the estimate relative to the original time series for every time step (bottom right).}
    \label{fig:fisherKPP}
\end{figure}

For the sake of simplicity all parameters (including the time delay $\tau$) are set to $1.0$, boundary conditions are periodic and the initial condition $u_0(x) = \text{sech}^2\left(0.2 \left(x - \frac{L_x}{2}\right)\right)$ is set for all $t\leq 0$. ($L_x$ is the domain size, here $L_x=90$.) A space-time plot of the time simulation is shown figure~\ref{fig:fisherKPP}\,(a).

To account for the time-delayed variable, we add 7 terms in $u(t-\tau)$ to our default library with $m=3$ and $p=4$ and add $\tau$ to the hyperparameters for optimization, such that some library terms now change during optimization.
This results in the identified model:
\begin{equation}\label{eqn:fisher_kpp_est}
    \partial_t \hat{u}(t) = 1.02 \partial_{x}^2 \hat{u}(t) + 0.98\hat{u}(t)  - 0.97u(t)\hat{u}(t-1.02)
\end{equation}
(see  Appendix~\ref{apx:results} table~\ref{tab:apx_fkpp} for complete results).
Again we find that the correct model is identified out of 23 candidate terms for the right-hand side. Figure~\ref{fig:fisherKPP}\,(b) shows a space-time plot of the estimated model. While we see that the error of the estimation is greatest at the propagation fronts (see figure~\ref{fig:fisherKPP}\,(c)) and increases over time (see figure~\ref{fig:fisherKPP}\,(d)), the average deviation of the reconstructed time series seems to approach a constant value (see again figure~\ref{fig:fisherKPP}\,(d)).

\newpage

\section{Conclusion and Outlook}
\label{sec:outlook}
In this work, we have shown how Bayesian optimization can improve data-driven model estimation methods, by automatically finding hyperparameters of the numerical method and/or of the aimed-at model. This has allowed us to extend the thresholding algorithm used in reference~\cite{Brunton2016} with additional thresholds, which affect individual parts of the estimated model structure. This has enabled model recovery from synthetic data with parameters of different orders of magnitude including models that show relaxational, time-periodic and chaotic behaviors.
We have extended the method introduced in reference~\cite{Kroll2025} for ODE models toward PDE and delay-PDE models.

In sections~\ref{sssec:baseline} and~\ref{sssec:space_steps_and_noise} we have shown that, for an elementary case, the overall performance and robustness against under-sampling in space and against the occurrence of additive noise are similar for our method implemented in TSME and the established method implemented in PySINDy. For both methods, further considerations are required when dealing with any significant amount of noise. In our case, this implies that the supplied data need to be de-noised before estimation. In the future, one may adapt more robust variations of the SINDy algorithm (such as, the one presented in reference.~\cite{Fasel2022, Maddu2022}) for use with hyperparameter optimization and the BIC.

While TSME results in higher computational cost, we have shown that it outperforms PySINDy in cases of data that are more sparsely sampled in time (section~\ref{sssec:var_sample}). TSME  allows for a wider range of permissible model structures and provides an implementation that allows for user-defined library terms. Furthermore, section~\ref{ssec:allen_cahn} has shown that our method performs well in scenarios where conservation laws typically require additional constraints. In particular,  in section~\ref{ssec:cahn_hilliard} we have recovered a Cahn-Hilliard model where PySINDy's functional library is ill-suited. Notably, as long as the employed data represent dynamics that involves a range of field values that is not too small the Cahn-Hilliard and Allen-Cahn systems can still be identified even without reaching extremal values or an equilibrium, because the library functions naturally extrapolate toward the extremal values. Then, once a model has been estimated it may be simulated in time to any extent and thus can reach values not present in the initial data. Several points related to ill-posedness \cite{Brunk_2023} have been discussed in section~\ref{ssec:cahn_hilliard}.

Additionally, via the usage of additional hyperparameters our method can accommodate on the one hand differences in the order of magnitude of parameters either between equations for the different field variables or between groups of terms within an equation (section~\ref{ssec:fhn}) and on the other hand added system parameters such as a time delay (section~\ref{ssec:fisherkpp}). This would typically require either a grid-based search through the hyperparameter space or an altogether altered implementation compared to the one presented in reference~\cite{deSilva2020}. We have shown that our method is very flexible and can accommodate any number of hyperparameters. This can be used to accommodate semantic connections between ansatz functions by making them share an additional hyperparameter. This yields more precise results as shown for some cases in section~\ref{ssec:rd_chaotic}.

Since model estimation of higher quality takes precedent over computation time, our results imply that the use of time integration should in the future be incorporated into other existing methods. Our method can be further improved by including sub-sampling or gradient-based optimization. Its library can easily accommodate more complex ansatz functions, which rely on additional hyperparameters, such as external driving forces or temperature dependent functions. This greatly increases the number of permissible modeling approaches for unknown systems. Another topic of interest are models with discontinuous solutions, e.g., a thin-film equation with a wetting energy that does not diverge for zero film height. While the present method could apply to such problems in theory, they should be treated in a dedicated manner, either by adapting the target function of the optimization or the optimization scheme itself.

 Importantly, after the synthetic data used here, in the future, the method needs to be applied to data obtained in discrete stochastic model approaches and experiments, where the continuum description of the dynamics is not explicitly known (but ideally a rich theoretical background exists that allows one to tune the term library and hyperparameter strategy).

\bibliography{literatur}

@InBook{Ahmed2021,
  author    = {Ahmed, S. E. And San, O. and Lakshmivarahan, S.},
  booktitle = {Advances in Nonlinear Dynamics},
  title     = {Forward sensitivity analysis of the {F}itz{H}ugh--{N}agumo system: parameter estimation},
  doi       = {10.1007/978-3-030-81170-9_9},
  isbn      = {9783030811709},
  pages     = {93--103},
  publisher = {Springer International Publishing},
  journal   = {Nodycon Conference Proceedings Series},
  year      = {2021},
  address   = {Cham, Switzerland}
}

@Article{Bandeira2024,
  author    = {Bandeira, J. G. P. and Buske, D. and De Quadros, R. S. and Kurz, G. B.},
  title     = {{A}llen-{C}ahn equation for modeling temporal evolution of non-conserved field variables in cancer cell migration},
  doi       = {10.5902/2179460x87268},
  number    = {esp. 1},
  pages     = {e87268},
  volume    = {46},
  journal   = {Ci{\^{e}}ncia E Natura},
  publisher = {Universidade Federal de Santa Maria},
  year      = {2024},
}

@InProceedings{Bergstra2011,
  author    = {Bergstra, J. and Bardenet, R. and Bengio, Y. and K\'{e}gl, B.},
  booktitle = {Advances in Neural Information Processing Systems},
  title     = {Algorithms for Hyper-Parameter Optimization},
  editor    = {J. Shawe-Taylor and R. Zemel and P. Bartlett and F. Pereira and K.Q. Weinberger},
  publisher = {Curran Associates, Inc.},
  volume    = {24},
  year      = {2011},
}

@Article{Bergstra2015,
  author    = {Bergstra, J. and Komer, B. and Eliasmith, C. and Yamins, D. and Cox, D. D.},
  title     = {{Hyperopt}: a {P}ython library for model selection and hyperparameter optimization},
  doi       = {10.1088/1749-4699/8/1/014008},
  number    = {1},
  pages     = {014008},
  volume    = {8},
  journal   = {Computational Science \& Discovery},
  publisher = {IOP Publishing},
  year      = {2015},
}

@Article{Bock1984,
  author    = {Bock, H. G. And Plitt, K. J.},
  title     = {A multiple shooting algorithm for direct solution of optimal control problems},
  doi       = {10.1016/s1474-6670(17)61205-9},
  number    = {2},
  pages     = {1603--1608},
  volume    = {17},
  journal   = {{IFAC} Proceedings Volumes},
  publisher = {Elsevier BV},
  year      = {1984},
}

@Article{Br1999,
  author    = {B{\"{a}}r, M. and Hegger, R. and Kantz, H.},
  title     = {Fitting partial differential equations to space-time dynamics},
  doi       = {10.1103/physreve.59.337},
  number    = {1},
  pages     = {337--342},
  volume    = {59},
  journal   = {Physical Review E},
  publisher = {American Physical Society (APS)},
  year      = {1999},
}

@Article{Brunton2016,
  author    = {Brunton, S. L. And Proctor, J. L. And Kutz, J. N.},
  title     = {Discovering governing equations from data by sparse identification of nonlinear dynamical systems},
  doi       = {10.1073/pnas.1517384113},
  number    = {15},
  pages     = {3932--3937},
  volume    = {113},
  journal   = {Proceedings of the National Academy of Sciences},
  publisher = {Proceedings of the National Academy of Sciences},
  year      = {2016},
}

@Article{Cahn1994,
  author    = {Cahn, J. W. And Novick-Cohen, A.},
  title     = {Evolution equations for phase separation and ordering in binary alloys},
  doi       = {10.1007/bf02188691},
  number    = {3-4},
  pages     = {877--909},
  volume    = {76},
  journal   = {Journal of Statistical Physics},
  publisher = {Springer Science and Business Media LLC},
  year      = {1994},
}

@Article{CebrinLacasa2024,
  author    = {Cebri{\'{a}}n-Lacasa, D. and Parra-Rivas, P. and Ruiz-Reyn{\'{e}}s, D. and Gelens, L.},
  title     = {Six decades of the {F}itz{H}ugh--{N}agumo model: a guide through its spatio-temporal dynamics and influence across disciplines},
  doi       = {10.1016/j.physrep.2024.09.014},
  pages     = {1--39},
  volume    = {1096},
  journal   = {Physics Reports},
  publisher = {Elsevier BV},
  year      = {2024},
}

@Misc{ddefind2024,
  author    = {Stephany, R.},
  title     = {{DDE}-{F}ind: {L}earning Delay Differential Equations from Noisy, Limited Data},
  doi       = {10.48550/arxiv.2405.02661},
  copyright = {arXiv.org perpetual, non-exclusive license},
  publisher = {arXiv},
  year      = {2024},
}

@Article{deSilva2020,
  author    = {De Silva, B. and Champion, K. and Quade, M. and Loiseau, J. and Kutz, J. And Brunton, S.},
  title     = {{PySINDy}: a {P}ython package for the sparse identification of nonlinear dynamical systems from data},
  doi       = {10.21105/joss.02104},
  number    = {49},
  pages     = {2104},
  volume    = {5},
  journal   = {Journal of Open Source Software},
  publisher = {The Open Journal},
  year      = {2020},
}

@Article{Dong2015,
  author    = {Dong, X. and Wang, C.},
  title     = {Identification of the {F}itz{H}ugh--{N}agumo model dynamics via deterministic learning},
  doi       = {10.1142/s021812741550159x},
  number    = {12},
  pages     = {1550159},
  volume    = {25},
  journal   = {International Journal of Bifurcation and Chaos},
  publisher = {World Scientific Pub Co Pte Lt},
  year      = {2015},
}

@Article{Dong2022,
  author    = {Dong, X. and Bai, Y. and Lu, Y. and Fan, M.},
  title     = {An improved sparse identification of nonlinear dynamics with akaike information criterion and group sparsity},
  doi       = {10.1007/s11071-022-07875-9},
  number    = {2},
  pages     = {1485--1510},
  volume    = {111},
  journal   = {Nonlinear Dynamics},
  publisher = {Springer Science and Business Media LLC},
  year      = {2022},
}

@Article{Ducrot2014,
  author    = {Ducrot, A. and Nadin, G.},
  title     = {Asymptotic behaviour of travelling waves for the delayed {F}isher--{KPP} equation},
  doi       = {10.1016/j.jde.2014.01.033},
  number    = {9},
  pages     = {3115--3140},
  volume    = {256},
  journal   = {Journal of Differential Equations},
  publisher = {Elsevier BV},
  year      = {2014},
}

@Article{Egan2024,
  author    = {Egan, K. and Li, W. and Carvalho, R.},
  title     = {Automatically discovering ordinary differential equations from data with sparse regression},
  doi       = {10.1038/s42005-023-01516-2},
  number    = {1},
  volume    = {7},
  journal   = {Communications Physics},
  publisher = {Springer Science and Business Media LLC},
  year      = {2024},
}

@Article{Fasel2022,
  author    = {Fasel, U. And Kutz, J. N. And Brunton, B. W. And Brunton, S. L.},
  title     = {{Ensemble-SINDy}: robust sparse model discovery in the low-data, high-noise limit, with active learning and control},
  doi       = {10.1098/rspa.2021.0904},
  number    = {2260},
  volume    = {478},
  journal   = {Proceedings of the Royal Society A: Mathematical, Physical and Engineering Sciences},
  publisher = {The Royal Society},
  year      = {2022},
}

@Misc{findiff,
  author = {Baer, M.},
  title  = {{findiff} software package},
  note   = {\url{https://github.com/maroba/findiff}},
  key    = {findiff},
  year   = {2018},
}

@Article{Fischer1937,
  author    = {Fisher, R. A.},
  title     = {The wave of advance of advantageous genes},
  doi       = {10.1111/j.1469-1809.1937.tb02153.x},
  number    = {4},
  pages     = {355--369},
  volume    = {7},
  journal   = {Annals of Eugenics},
  publisher = {Wiley},
  year      = {1937},
}

@Article{FitzHugh1961,
  author    = {Fitz{H}ugh, R.},
  title     = {Impulses and physiological states in theoretical models of nerve membrane},
  doi       = {10.1016/s0006-3495(61)86902-6},
  number    = {6},
  pages     = {445--466},
  volume    = {1},
  journal   = {Biophysical Journal},
  publisher = {Elsevier BV},
  year      = {1961},
}

@Article{Fullana1997,
  author    = {Fullana, J. and Rossi, M. and Zaleski, S.},
  title     = {Parameter identification in noisy extended systems: a hydrodynamic case},
  doi       = {10.1016/s0167-2789(96)00286-2},
  number    = {1-4},
  pages     = {564--575},
  volume    = {103},
  journal   = {Physica D: Nonlinear Phenomena},
  publisher = {Elsevier BV},
  year      = {1997},
}

@Article{Guay1995,
  author       = {Guay, M. And Mclean, D. D.},
  date         = {1995-12},
  journaltitle = {Computers \& Chemical Engineering},
  title        = {Optimization and sensitivity analysis for multiresponse parameter estimation in systems of ordinary differential equations},
  doi          = {10.1016/0098-1354(94)00120-0},
  issn         = {0098-1354},
  number       = {12},
  pages        = {1271--1285},
  volume       = {19},
  journal      = {Computers \& Chemical Engineering},
  mmonth        = dec,
  publisher    = {Elsevier BV},
  year         = {1995},
}

@Book{Hairer1993,
  title     = {Solving ordinary differential equations {I}},
  author    = {Hairer, E and Norsett, S. P. and Wanner, G.},
  publisher = {Springer},
  series    = {Springer Series in Computational Mathematics},
  edition   = {2},
  mmonth     = {aug},
  year      = {1993},
  doi       = {https://doi.org/10.1007/978-3-540-78862-1},
  address   = {Berlin, Germany}
}

@Article{harris2020,
  author    = {Harris, C. R. And Millman, K. J. and Van Der Walt, S. J. And Gommers, R. and Virtanen, P. and Cournapeau, D. and Wieser, E. and Taylor, J. and Berg, S. and Smith, N. J. And Kern, R. and Picus, M. and Hoyer, S. and Van Kerkwijk, M. H. And Brett, M. and Haldane, A. and Del R{\'{i}}o, J. F. and Wiebe, M. and Peterson, P. and G{\'{e}}rard-Marchant, P. and Sheppard, K. and Reddy, T. and Weckesser, W. and Abbasi, H. and Gohlke, C. and Oliphant, T. E.},
  title     = {Array programming with {NumPy}},
  doi       = {10.1038/s41586-020-2649-2},
  number    = {7825},
  pages     = {357--362},
  volume    = {585},
  journal   = {Nature},
  publisher = {Springer Science and Business Media LLC},
  year      = {2020},
}

@Article{Hodgkin1952,
  author    = {Hodgkin, A. L. And Huxley, A. F.},
  title     = {A quantitative description of membrane current and its application to conduction and excitation in nerve},
  doi       = {10.1113/jphysiol.1952.sp004764},
  number    = {4},
  pages     = {500--544},
  volume    = {117},
  journal   = {The Journal of Physiology},
  publisher = {Wiley},
  year      = {1952},
}

@InBook{Houska2012,
  author    = {Houska, B. and Logist, F. and Diehl, M. and Impe, J. V.},
  booktitle = {Identification for Automotive Systems},
  date      = {2012},
  title     = {A tutorial on numerical methods for state and parameter estimation in nonlinear dynamic systems},
  doi       = {10.1007/978-1-4471-2221-0_5},
  isbn      = {9781447122210},
  pages     = {67--88},
  publisher = {Springer London},
  issn      = {1610-7411},
  journal   = {Lecture Notes in Control and Information Sciences},
  year      = {2012},
  address   = {London, England}
}

@Article{Hussain2019,
  author    = {Hussain, S. and Shah, A. and Ayub, S. and Ullah, A.},
  title     = {An approximate analytical solution of the {A}llen-{C}ahn equation using homotopy perturbation method and homotopy analysis method},
  doi       = {10.1016/j.heliyon.2019.e03060},
  number    = {12},
  pages     = {e03060},
  volume    = {5},
  journal   = {Heliyon},
  publisher = {Elsevier BV},
  year      = {2019},
}

@Article{Kim2016,
  author    = {Kim, J. and Lee, S. and Choi, Y. and Lee, S. and Jeong, D.},
  title     = {Basic principles and practical applications of the {C}ahn--{H}illiard equation},
  doi       = {10.1155/2016/9532608},
  pages     = {1--11},
  volume    = {2016},
  journal   = {Mathematical Problems in Engineering},
  publisher = {Wiley},
  year      = {2016},
}

@Article{Koch2021,
  author    = {Koch, J.},
  title     = {Data-driven modeling of nonlinear traveling waves},
  doi       = {10.1063/5.0043255},
  number    = {4},
  volume    = {31},
  journal   = {Chaos: An Interdisciplinary Journal of Nonlinear Science},
  publisher = {AIP Publishing},
  year      = {2021},
}

@Article{kolmogorov1937,
  author  = {Kolmogorov, A. And Petrovskii, I. And Piscounov, N.},
  title   = {A study of the diffusion equation with increase in the amount of substance, and its application to a biological problem. {V}{M} {T}ikhomirov editor},
  volume  = {1},
  journal = {Bull. Moscow Univ. Math. Mech},
  year    = {1937},
}

@Misc{Kroll2025,
  author        = {Kroll, T. W. And Kamps, O.},
  title         = {Sparse identification of evolution equations via {B}ayesian model selection},
  doi           = {10.48550/arxiv.2501.01476},
  eprint        = {2501.01476},
  eprinttype    = {arxiv},
  archiveprefix = {arXiv},
  primaryclass  = {physics.data-an},
  year          = {2025},
}

@book{Ljung1998,
  address = {Philadelphia, PA},
  author = {Ljung, L.},
  edition = {2},
  language = {en},
  publisher = {Prentice Hall},
  series = {Prentice Hall information and system sciences series},
  title = {{S}ystem identification: {T}heory for the user},
  year = {1998}
}

@Article{Long2019,
  author    = {Long, Z. and Lu, Y. and Dong, B.},
  title     = {{PDE-Net} 2.0: learning {PDE}s from data with a numeric-symbolic hybrid deep network},
  doi       = {10.1016/j.jcp.2019.108925},
  pages     = {108925},
  volume    = {399},
  journal   = {Journal of Computational Physics},
  publisher = {Elsevier BV},
  year      = {2019},
}

@Misc{Machlanski2023,
  author    = {Machlanski, D. And Samothrakis, S. And Clarke, P.},
  title     = {{H}yperparameter {T}uning and {M}odel {E}valuation in {C}ausal {E}ffect {E}stimation},
  doi       = {10.48550/arxiv.2303.01412},
  copyright = {Creative Commons Attribution 4.0 International},
  publisher = {arXiv},
  year      = {2023},
}

@Article{Maddu2022,
  author    = {Maddu, S. and Cheeseman, B. L. And Sbalzarini, I. F. And M{\"{u}}ller, C. L.},
  title     = {Stability selection enables robust learning of differential equations from limited noisy data},
  doi       = {10.1098/rspa.2021.0916},
  number    = {2262},
  volume    = {478},
  journal   = {Proceedings of the Royal Society A: Mathematical, Physical and Engineering Sciences},
  publisher = {The Royal Society},
  year      = {2022},
}

@Article{Mangan2017,
  author    = {Mangan, N. M. And Kutz, J. N. And Brunton, S. L. And Proctor, J. L.},
  title     = {Model selection for dynamical systems via sparse regression and information criteria},
  doi       = {10.1098/rspa.2017.0009},
  number    = {2204},
  pages     = {20170009},
  volume    = {473},
  journal   = {Proceedings of the Royal Society A: Mathematical, Physical and Engineering Sciences},
  publisher = {The Royal Society},
  year      = {2017},
}

@Article{Messenger2021,
  author    = {Messenger, D. A. And Bortz, D. M.},
  title     = {Weak {SINDy}: {G}alerkin-based data-driven model selection},
  doi       = {10.1137/20{m}1343166},
  number    = {3},
  pages     = {1474--1497},
  volume    = {19},
  journal   = {Multiscale Modeling \& Simulation},
  publisher = {Society for Industrial & Applied Mathematics (SIAM)},
  year      = {2021},
}

@Article{Nagumo1962,
  author    = {Nagumo, J. And Arimoto, S. And Yoshizawa, S.},
  title     = {An active pulse transmission line simulating nerve axon},
  doi       = {10.1109/jrproc.1962.288235},
  number    = {10},
  pages     = {2061--2070},
  volume    = {50},
  journal   = {Proceedings of the {IRE}},
  publisher = {Institute of Electrical and Electronics Engineers (IEEE)},
  year      = {1962},
}

@Article{Naozuka2022,
  author    = {Naozuka, G. T. And Rocha, H. L. And Silva, R. S. And Almeida, R. C.},
  title     = {{SINDy-SA} framework: enhancing nonlinear system identification with sensitivity analysis},
  doi       = {10.1007/s11071-022-07755-2},
  number    = {3},
  pages     = {2589--2609},
  volume    = {110},
  journal   = {Nonlinear Dynamics},
  publisher = {Springer Science and Business Media LLC},
  year      = {2022},
}

@Article{Omejc2024,
  author    = {Omejc, N. and Gec, B. and Brence, J. and Todorovski, L. and D{\v{z}}eroski, S.},
  title     = {Probabilistic grammars for modeling dynamical systems from coarse, noisy, and partial data},
  doi       = {10.1007/s10994-024-06522-1},
  number    = {10},
  pages     = {7689--7721},
  volume    = {113},
  journal   = {Machine Learning},
  publisher = {Springer Science and Business Media LLC},
  year      = {2024},
}

@Article{Rudy2017,
  author    = {Rudy, S. H. And Brunton, S. L. And Proctor, J. L. And Kutz, J. N.},
  title     = {Data-driven discovery of partial differential equations},
  doi       = {10.1126/sciadv.1602614},
  number    = {4},
  volume    = {3},
  journal   = {Science Advances},
  publisher = {American Association for the Advancement of Science (AAAS)},
  year      = {2017},
}

@Article{Rudy2019,
  author    = {Rudy, S. and Alla, A. and Brunton, S. L. And Kutz, J. N.},
  title     = {Data-driven identification of parametric partial differential equations},
  doi       = {10.1137/18{m}1191944},
  number    = {2},
  pages     = {643--660},
  volume    = {18},
  journal   = {{SIAM} Journal on Applied Dynamical Systems},
  publisher = {Society for Industrial & Applied Mathematics (SIAM)},
  year      = {2019},
}

@Article{Sandoz2023,
  author    = {Sandoz, A. and Ducret, V. and Gottwald, G. A. And Vilmart, G. and Perron, K.},
  title     = {{SINDy} for delay-differential equations: application to model bacterial zinc response},
  doi       = {10.1098/rspa.2022.0556},
  number    = {2269},
  volume    = {479},
  journal   = {Proceedings of the Royal Society A: Mathematical, Physical and Engineering Sciences},
  publisher = {The Royal Society},
  year      = {2023},
}

@Article{SciPy2020,
  author    = {Virtanen, P. and Gommers, R. and Oliphant, T. E. And Haberland, M. and Reddy, T. and Cournapeau, D. and Burovski, E. and Peterson, P. and Weckesser, W. and Bright, J. and Van Der Walt, S. J. And Brett, M. and Wilson, J. and Millman, K. J. and Mayorov, N. and Nelson, A. R. J. And Jones, Eric and Kern, Robert and Larson, Eric and Carey, C. J. And Polat, {{\.{I}}}lhan and Feng, Yu and Moore, Eric W. And Vanderplas, Jake and Laxalde, D. and Perktold, J. and Cimrman, R. and Henriksen, I. and Quintero, E. A. And Harris, C. R. And Archibald, A. M. And Ribeiro, A. H. And Pedregosa, F. and Van Mulbregt, P. and Vijaykumar, A. and Bardelli, A. P. and Rothberg, A. and Hilboll, A. and Kloeckner, A. and Scopatz, A. and Lee, A. and Rokem, A. and Woods, C. N. and Fulton, C. and Masson, C. and H{\"{a}}ggstr{\"{o}}m, C. and Fitzgerald, C. and Nicholson, D. A. And Hagen, D. R. And Pasechnik, D. V. And Olivetti, E. and Martin, E. and Wieser, E. and Silva, F. and Lenders, F. and Wilhelm, F. and Young, G. And Price, G. A. And Ingold, G. and Allen, G. E. And Lee, G. R. And Audren, H. and Probst, I. and Dietrich, J. P. And Silterra, J. and Webber, J. T. And Slavi{\v{c}}, J. and Nothman, J. and Buchner, J. and Kulick, J. and Sch{\"{o}}nberger, J. L. And De Miranda Cardoso, J. V. and Reimer, J. and Harrington, J. and Rodr{\'{i}}guez, J. L. C. and Nunez-Iglesias, J. and Kuczynski, J. and Tritz, K. and Thoma, M. and Newville, M. and K{\"{u}}mmerer, M. and Bolingbroke, M. and Tartre, M. and Pak, M. and Smith, N. J. And Nowaczyk, N. and Shebanov, N. and Pavlyk, O. and Brodtkorb, P. A. And Lee, P. and Mcgibbon, R. T. And Feldbauer, R. and Lewis, S. and Tygier, S. and Sievert, S. and Vigna, S. and Peterson, S. and More},
  title     = {{SciPy} 1.0: fundamental algorithms for scientific computing in {P}ython},
  doi       = {10.1038/s41592-019-0686-2},
  number    = {3},
  pages     = {261--272},
  volume    = {17},
  adsurl    = {https://rdcu.be/b08{Wh}},
  journal   = {Nature Methods},
  publisher = {Springer Science and Business Media LLC},
  year      = {2020},
}

@Article{Shraiman1992,
  author    = {Shraiman, B. I. And Pumir, A. And Van Saarloos, W. And Hohenberg, P. C. And Chat{\'{e}}, H. And Holen, M.},
  title     = {Spatiotemporal chaos in the one-dimensional complex {G}inzburg-{L}andau equation},
  doi       = {10.1016/0167-2789(92)90001-4},
  number    = {3-4},
  pages     = {241--248},
  volume    = {57},
  journal   = {Physica D: Nonlinear Phenomena},
  publisher = {Elsevier BV},
  year      = {1992},
}

@Article{Stegemerten2020,
  author    = {Stegemerten, F. and Gurevich, S. V. And Thiele, U.},
  title     = {Bifurcations of front motion in passive and active {A}llen--{C}ahn-type equations},
  doi       = {10.1063/5.0003271},
  number    = {5},
  volume    = {30},
  journal   = {Chaos: An Interdisciplinary Journal of Nonlinear Science},
  publisher = {AIP Publishing},
  year      = {2020},
}

@Article{Tang2023,
  author    = {Tang, M. and Liao, W. and Kuske, R. and Kang, S. H.},
  title     = {Weakident: weak formulation for identifying differential equation using narrow-fit and trimming},
  doi       = {10.1016/j.jcp.2023.112069},
  pages     = {112069},
  volume    = {483},
  journal   = {Journal of Computational Physics},
  publisher = {Elsevier BV},
  year      = {2023},
}

@online{TSME,
  author = {Mai, O.},
  note = {\url{https://pypi.org/project/tsme/}},
  title = {{TSME} {P}y{P}i repository},
  url = {https://pypi.org/project/tsme/},
  urldate = {2025-07-24},
  year = {2024}
}

@online{TSMEdocumentationCHE,
  author = {Mai, O.},
  note = {\url{https://nonlinear-physics.zivgitlabpages.uni-muenster.de/ag-kamps/tsme/source/notebooks/cahn_hilliard_equation.html}},
  title = {{TSME} documentation: {C}ahn-{H}illiard {E}quation},
  url = {https://nonlinear-physics.zivgitlabpages.uni-muenster.de/ag-kamps/tsme/source/notebooks/cahn_hilliard_equation.html},
  urldate = {2025-07-24},
  year = {2022}
}

@online{TSMEgithub,
  author = {Mai, O.},
  note = {\url{https://github.com/CeNoS-CoSy/tsme_examples}},
  title = {{TSME} examples {G}it{H}ub repository},
  url = {https://github.com/CeNoS-CoSy/tsme_examples},
  urldate = {2025-07-24},
  year = {2025}
}

@Article{Voss1998,
  author    = {Voss, H. And B{\"{u}}nner, M. And Abel, M.},
  title     = {Identification of continuous, spatiotemporal systems},
  doi       = {10.1103/physreve.57.2820},
  number    = {3},
  pages     = {2820--2823},
  volume    = {57},
  journal   = {Physical Review E},
  publisher = {American Physical Society (APS)},
  year      = {1998},
}

@Article{Voss1999,
  author    = {Voss, H. U. And Kolodner, P. and Abel, M. and Kurths, J.},
  title     = {Amplitude equations from spatiotemporal binary-fluid convection data},
  doi       = {10.1103/physrevlett.83.3422},
  number    = {17},
  pages     = {3422--3425},
  volume    = {83},
  journal   = {Physical Review Letters},
  publisher = {American Physical Society (APS)},
  year      = {1999},
}

@Article{VOSS2004,
  author    = {Voss, H. U. And Timmer, J. and Kurths, J.},
  title     = {Nonlinear dynamical system identification from uncertain and indirect measurements},
  doi       = {10.1142/s0218127404010345},
  number    = {06},
  pages     = {1905--1933},
  volume    = {14},
  journal   = {International Journal of Bifurcation and Chaos},
  publisher = {World Scientific Pub Co Pte Ltd},
  year      = {2004},
}

@Article{Wan2021,
  author    = {Wan, C. and Yu, Z. and Wang, F. and Liu, X. and Li, J.},
  title     = {Experiment data-driven modeling of tokamak discharge in {EAST}},
  doi       = {10.1088/1741-4326/abf419},
  number    = {6},
  pages     = {066015},
  volume    = {61},
  journal   = {Nuclear Fusion},
  publisher = {IOP Publishing},
  year      = {2021},
}

@Article{Wang2019,
  author    = {Wang, Z. And Huan, X. And Garikipati, K.},
  title     = {Variational system identification of the partial differential equations governing the physics of pattern-formation: inference under varying fidelity and noise},
  doi       = {10.1016/j.cma.2019.07.007},
  pages     = {44--74},
  volume    = {356},
  journal   = {Computer Methods in Applied Mechanics and Engineering},
  publisher = {Elsevier BV},
  year      = {2019},
}

@Article{Zhang2025,
  author    = {Zhang, J. and Hu, H. and Huang, C.},
  title     = {Wave fronts for a class of delayed {F}isher--{KPP} equations},
  doi       = {10.1016/j.aml.2024.109406},
  pages     = {109406},
  volume    = {163},
  journal   = {Applied Mathematics Letters},
  publisher = {Elsevier BV},
  year      = {2025},
}

@InBook{Zhao2018,
  author    = {Zhao, J. and Wang, W. and Sheng, C.},
  booktitle = {Data-Driven Prediction for Industrial Processes and Their Applications},
  title     = {Parameter estimation and optimization},
  doi       = {10.1007/978-3-319-94051-9_7},
  isbn      = {9783319940519},
  pages     = {269--350},
  publisher = {Springer International Publishing},
  journal   = {Information Fusion and Data Science},
  year      = {2018},
  address   = {Basel, Switzerland}
}

@Article{Zhao2020,
  author    = {Zhao, H. and Storey, B. D. And Braatz, R. D. And Bazant, M. Z.},
  title     = {Learning the physics of pattern formation from images},
  doi       = {10.1103/physrevlett.124.060201},
  number    = {6},
  volume    = {124},
  journal   = {Physical Review Letters},
  publisher = {American Physical Society (APS)},
  year      = {2020},
}

@Article{Zheng2015,
  author    = {Zheng, Q. and Shen, J.},
  title     = {Pattern formation in the {F}itz{H}ugh--{N}agumo model},
  doi       = {10.1016/j.camwa.2015.06.031},
  number    = {5},
  pages     = {1082--1097},
  volume    = {70},
  journal   = {Computers \& Mathematics with Applications},
  publisher = {Elsevier BV},
  year      = {2015},
}

@Misc{zubov2021neuralpde,
  author     = {Zubov,  K. and McCarthy,  Z. and Ma,  Y. and Calisto,  F. and Pagliarino,  V. and Azeglio,  S. and Bottero,  L. and Luj\'{a}n,  E. and Sulzer,  Va. and Bharambe,  A. and Vinchhi,  N. and Balakrishnan,  K. and Upadhyay,  D. and Rackauckas,  C.},
  title      = {{NeuralPDE}: Automating Physics-Informed Neural Networks ({PINN}s) with Error Approximations},
  eprint     = {2107.09443},
  eprinttype = {arxiv},
  year       = {2021},
}

@Article{Brunk_2023,
  author    = {Brunk, A. and Egger, H. and Habrich, O.},
  title     = {On uniqueness and stable estimation of multiple parameters in the {C}ahn--{H}illiard equation},
  doi       = {10.1088/1361-6420/acca44},
  number    = {6},
  pages     = {065002},
  volume    = {39},
  journal   = {Inverse Problems},
  mmonth     = apr,
  publisher = {IOP Publishing},
  year      = {2023},
}

@Article{CrMa2009rmp,
  author   = {Craster, R. V. And Matar, O. K.},
  title    = {Dynamics and stability of thin liquid films},
  doi      = {10.1103/RevModPhys.81.1131},
  pages    = {1131--1198},
  volume   = {81},
  journal  = {Rev. Mod. Phys.},
  uwekeys  = {Surfactant__, Surfactant_equation-of-state__, Review__, Dewetting__, Capillarity__, Thin_film_liquid__, Marangoni_flow__, Long_wave_instability__, Lubrication_approximation__,Thin_film_two_layer__},
  year     = {2009},
}

\FloatBarrier
\appendix

\section{Generating the library of functions}\label{apx:generating_library}
In this section we detail different strategies to generate the set of library functions used by the method developed in section~\ref{sec:num_methods}. They are based on the physical quantity $u(x, t)\in \mathbb{R}^n$ we ultimately wish to describe, with $t$ being the time and $x\in\mathbb{R}^d, d\in\{1, 2, 3\}$ the position in $d$-dimensional space. Each library function has a coefficient associated with it, that will be optimized to best describe data of $u$ by the method of section~\ref{sec:num_methods}.
\subsection{Default Library}\label{apx:default_library}
We construct a set of monomials (or power products) in the $n$ field variables $u_i$ from zeroth order up to order $m$. Then, for each non-trivial monomial we compute all (partial) spatial derivatives up to order $p$. More specifically let $\boldsymbol{\alpha} = (\alpha_1, \alpha_2, \ldots, \alpha_n)$ be a multi-index such that
\begin{equation}
    \boldsymbol{\alpha} \in \mathbb{N}_0^n, |\boldsymbol{\alpha}| = \sum_{i=1}^n \alpha_i \leq m,
\end{equation}
then, each monomial is of the form
\begin{equation}
    U_{\boldsymbol{\alpha}} = \prod_{i=1}^{n} u_i^{\alpha} = u_1^{\alpha_1}u_2^{\alpha_2} \cdots u_n^{\alpha_n}.
\end{equation}
For example, in the case $n=2$, $m=2$ and $p=0$ the set of monomials is
\begin{equation*}
    \left\{1, u_1,u_2, u_1^2, u_1 u_2, u_2^2 \right\}.
\end{equation*}
Similarly for the spatial derivatives, let $\boldsymbol{\beta} = (\beta_1, \beta_2, \ldots, \beta_d)$, where $d$ is the spatial dimension, be a multi-index such that
\begin{equation}
    \boldsymbol{\beta} \in \mathbb{N}_0^d, |\boldsymbol{\beta}| = \sum_{j=1}^d \beta_j \leq p,
\end{equation}
where the corresponding differential operator is
\begin{equation}
    \partial_x^{\boldsymbol{\beta}} = \partial_{x_1}^{\beta_1} \partial_{x_2}^{\beta_2} \cdots \partial_{x_d}^{\beta_d}.
\end{equation}
For example, with $d=2$ and $p=2$ the set of differential operators is
\begin{equation*}
    \left\{ \partial_{x_1}^0 \partial_{x_2}^0, \partial_{x_1}^1 \partial_{x_2}^0, \partial_{x_1}^0 \partial_{x_2}^1, \partial_{x_1}^2 \partial_{x_2}^0, \partial_{x_1}^1 \partial_{x_2}^1, \partial_{x_1}^0 \partial_{x_2}^2\right\}.
\end{equation*}
Now we apply each differential operator to each monomial except the one for $|\alpha| = 0$. For the sake of sequencing, let us group all terms by order $|\boldsymbol{\beta}|$. This gives the vector of ansatz functions, by order $q$ of their respective differential operator:
\begin{equation}
    \Theta^{(q)} = \left[ \partial_x^{\boldsymbol{\beta}} U_{\boldsymbol{\alpha}} ~ \big|~ 1\leq |\boldsymbol{\alpha}| \leq m, |\boldsymbol{\beta}| = q \right].
\end{equation}
So for any maximum order $p$, we take all ansatz functions with $0 \leq q \leq p$.
Combining the earlier examples ($d=2, n=2, m=2$, and $p=2$, i.e., $q=0, 1, 2$):
\begin{align*}
    \Theta^{(0)} &= \left[u_1,u_2, u_1^2, u_1 u_2, u_2^2  \right] \\
    \Theta^{(1)} &= [\partial_{x_1} u_1, \partial_{x_1} u_2, \partial_{x_1} u_1^2, \partial_{x_1} (u_1 u_2), \partial_{x_1} u_2^2, \partial_{x_2} u_1, \partial_{x_2} u_2, \partial_{x_2} u_1^2, \partial_{x_2} (u_1 u_2), \partial_{x_2} u_2^2] \\
    \Theta^{(2)} &= \left[ \partial_{x_1}^2 u_1, \partial_{x_1}^2 u_2, \partial_{x_1}^2 u_1^2, \ldots, \partial_{x_1} \partial_{x_2} u_1, \partial_{x_1} \partial_{x_2} u_2, \ldots, \partial_{x_2}^2 u_2^2 \right].
\end{align*}
We note here, that terms such as $\partial_x (u_1 u_2)$ in $\Theta^{(1)}$ could be expressed as two terms ($u_1\partial_x u_2$ and $u_2\partial_x u_1$) and therefore could have different coefficients within the estimation procedure, e.g., for the case of nonlinear diffusion. An entirely inclusive case could be based on the generalized product rule.
So finally our entire library is comprised the concatenation of all the elements of these vectors:
\begin{equation}
    \Theta = [1\, \Theta^{(0)}\,  \Theta^{(1)}  \ldots \, \Theta^{(p)}]^T.
\end{equation}
The total number $k$ of all these elements may become fairly large due to combinatorics:
\begin{align}
    k &= 1 +\left( \text{Number of monomials with }|\boldsymbol{\alpha}|\geq1 \right)\, \times  \left(\text{Number of derivatives}\right) \notag \\
      &= 1 +  \left( \sum_{r=1}^m \binom{n + r - 1}{r} \right) \cdot \left( \sum_{q=0}^p \binom{d + q - 1}{q} \right).
\end{align}
Following through with our example for $d=2, n=2, m=2, p=2$ we find $k=31$. This default library may be modified by removing certain terms or adding terms, such as functions with explicit time dependence or $u(x, t-\tau)$, where $\tau$ is a set time delay. Also non-analytic functions may be included.

\subsection{Alternative Library: PySINDy}\label{apx:ssec_library_sindy}
For the sake of comparison with reference~\cite{deSilva2020} we recount the library used in PySINDy. It consists of power products as in our case and terms linear in spatial derivatives of $u$. 
This leads to the following entries in the library:
\begin{equation}
	    \Theta^{(q)} = \left[ U_{\boldsymbol{\alpha}}\partial_x^{\boldsymbol{\beta}} u_i ~ \big|~ 0\leq |\boldsymbol{\alpha}| \leq m, |\boldsymbol{\beta}| = q, 0 \leq i \leq n \right]
\end{equation}
and just as before (only now with the trivial element included in $\Theta^{(0)}$):
\begin{equation*}
	\Theta = [\Theta^{(0)} \, \Theta^{(1)} \, \ldots \, \Theta^{(p)}]^T.
\end{equation*}

For example with $d=2, n=2, m=2, p=2$:
\begin{align*}
    \Theta^{(0)} &= \left[1, u_1,u_2, u_1^2, u_1 u_2, u_2^2  \right] \\
    \Theta^{(1)} &= [\partial_{x_1} u_1, \partial_{x_1} u_2, 
    u_1\partial_{x_1} u_1, u_2\partial_{x_1} u_1, u_1\partial_{x_1} u_2, u_2\partial_{x_1} u_2, \\
	&~~~~~~~u_1^2\partial_{x_1} u_1, u_1 u_2\partial_{x_1} u_1, u_1 u_2\partial_{x_1} u_2, u_2^2 \partial_{x_1} u_1, u_2^2\partial_{x_1} u_2, \\
	&~~~~~~~ \partial_{x_2} u_1, \partial_{x_2} u_2, \ldots, u_2^2\partial_{x_2} u_2] \\
    \Theta^{(2)} &= [ \partial_{x_1}^2 u_1, \partial_{x_1}^2 u_2, u_1\partial_{x_1}^2 u_1, \ldots, u_2^2 \partial_{x_1} u_2, \\ 
    &~~~~~~~ \partial_{x_1} \partial_{x_2} u_1,\partial_{x_1} \partial_{x_2} u_2, u_1 \partial_{x_1} \partial_{x_2} u_1, \ldots, u_2^2\partial_{x_1} \partial_{x_2} u_2, \\
    &~~~~~~~ \partial_{x_2}^2 u_1,\partial_{x_2}^2 u_2 , u_1 \partial_{x_2}^2 u_1, \ldots, u_2^2 \partial_{x_2}^2 u_2 ]
\end{align*}
(Please note that here the ordering differs reference~\cite{deSilva2020}.)
The number $k$ of library elements in PySINDy is:
\begin{align}
    k &=\left( \text{Number of monomials} \right)\, \times   \left(1+\text{Number of derivatives}\right) \notag \\
      &= \binom{n + m}{m} \left(1 + n \sum_{q=1}^p \binom{d + q - 1}{q} \right),
\end{align}
which for this example gives $k=66$. Please note that in this case $m$ strictly only gives the highest order of monomial, not the highest order of the entire library (which is $m+1$ as each monomial is being multiplied by a derivative of a linear term in $u_i$). This approach gives more individual terms for the same values of parameters $d, n, m$ and $p$ than our approach, but not an identical order or coverage of the function space. 
\newpage
\section{Pseudo-Code for Estimation Procedure}\label{apx:pseudo_code}
\begin{algorithm}
\caption{Structure of the estimation method}
\label{alg:TSME}
\begin{algorithmic}
\State Load time series data $u$, time stamps $t$ and spacial dimensions $L_1,\ldots, L_d$
\If{delay $\tau$ is to be used}
    \State Ensure $u$ has values for $-\tau \leq t \leq 0$ or set those values to $u(t=0)$
\EndIf
\State Compute $\partial_t u$ via finite differences and time stamps $t$
\State Create a semantic representation of (default) library terms $\Theta$
\State (Optional) Modify library by removing or adding terms (such as $u(t-\tau)$)
\State Evaluate all library terms $\Theta$ at all time steps using FFT (or finite difference) for spacial derivatives, with applicable spatial domain sizes $L_1,\ldots, L_d$
\State Define search space of the TPE for all hyperparameters: thresholds $h_i$ and $\tau$ if applicable
\For{$l=0$ to $l=l_{\text{max}}$}
    \State Recompute any library terms that may have changed (e.g., time delayed terms)
    \State Perform sequential thresholding least-squares with $h_i$ to find values for $\hat{\sigma}$ minimizing $||\partial_t u - \hat{\sigma} \cdot \Theta ||_2$ 
    \State Perform time integration using $\hat{\sigma}$ (and $\tau$) to obtain $\hat{u}$
    \State Evaluate $\text{BIC}= s \ln (N_t) - 2 \ln (||u-\hat{u}||_2)$
    \State Use TPE w.r.t. BIC to find new values for $h_i$ (and $\tau$)
\EndFor

\end{algorithmic}
\end{algorithm}

\newpage
\section{Tutorials}\label{apx:tutorials}
\subsection{Cahn-Hilliard Equation}\label{apx:che_tutorial}
This section is a modified example taken from the documentation of the Python package TSME found in reference~\cite{TSMEdocumentationCHE} or \href{https://nonlinear-physics.zivgitlabpages.uni-muenster.de/ag-kamps/tsme/source/notebooks/cahn_hilliard_equation.html}{here}. and will walk through the Python code needed to reproduce the results shown here.

We consider the Cahn-Hilliard equation in 2D ($d=2$, $n=1$) discussed in section~\ref{ssec:cahn_hilliard}:
\begin{equation*}
    \partial_t u = \nabla^2 \left( u^3 - u - \nabla^2 u \right).
\end{equation*}
We begin by performing a time simulation. As we are dealing with a spatially extended 2D system we first set our spatial discretization \texttt{N} and domain sizes \texttt{Lx} and \texttt{Ly} in $x$ and $y$ direction.  Then, we define a time interval \texttt{time\_interval}, as well as the time stamps \texttt{t\_eval} where we want to sample the trajectory. The initial condition \texttt{u0} is set to uniform random values over the interval $[-0.25, 0.25)$.

\begin{pyin}[labelOfTheSecondInput]
import numpy as np
np.random.seed(12389)

N = 128
Lx = 90
Ly = 90
domain = ((0, Lx), (0, Ly))
time_interval = [0, 15]
t_eval = np.linspace(time_interval[0], time_interval[-1], 100)
u0 = (np.random.random((N, N)) - 0.5) * 0.5

\end{pyin}

Next, we create the Cahn-Hilliard problem using TSME as an implementation of the supplied \texttt{AbstractTimeSimulator} class:

\begin{pyin}
from tsme.time_simulation import AbstractTimeSimulator

class CHESim(AbstractTimeSimulator):
    def __init__(self, ic=None, dom=domain, params=None, 
                 bc="periodic", diff="finite_difference"):
        super().__init__(ic, domain=dom, bc=bc, diff=diff)
        if params is None:
            params = [1.0]
        self.a = params[0]
    
    def rhs(self, t, u_in):
        u = u_in[0]
        
        f = u ** 3 - u \
        - self.a * (self.diff.d_dx(u, 2) + self.diff.d_dy(u, 2))
        
        u_next = self.diff.d_dx(f, 2) + self.diff.d_dy(f, 2)
    
        return np.array([u_next])
\end{pyin}
This class allows for a number of different expressions in the right-hand side function \texttt{rhs}. Here we use the spatial derivatives in both spatial directions to construct our Laplace operator.

Now all our preparations are done, and we can perform the actual time simulation:
\begin{pyin}
sim = CHESim(ic=np.array([u0]), dom=domain, params=[1.0])
sol = sim.simulate(time_interval, method="DOP853", t_eval=t_eval)

\end{pyin}

\begin{pyprint}
IVP: 100%|##########| 15.0/15.0 [06:19<00:00, 25.33s/ut]
\end{pyprint}
The output shown here is a progress bar (that would be updated in real time), which shows in percent how far in time the solver has progressed, how much time has elapsed since the computation began and how long the rest of the computation is estimated to take. Since the computation here is already finished, we see that it took about six and a half minutes to complete and that there is no computation time remaining.

Now we import the model that will estimate the differential equations.
\begin{pyin}
    from tsme.model_estimation.model import Model
\end{pyin}
All we need to do now is pass along the trajectory, domain size and the time stamps to our model. We can then automatically generate a library of possible test functions for our right-hand side. Here we take the powers of up to third order ($m=3$) and then, all spatial derivatives up to fourth order ($p=4$) of said powers (these include both spatial directions and their mixed derivatives). In total these are 46 terms.
\begin{pyin}
estimated_model = Model(sol, sim.time, phys_domain=domain)
estimated_model.init_library(3, 4)
estimated_model.print_library()
\end{pyin}

\begin{pyprint}
|   Index | Term                          |   Value 0 |
|---------|-------------------------------|-----------|
|       0 | 1.0                           |         0 |
|       1 | u[0]                          |         0 |
|       2 | u[0]*u[0]                     |         0 |
|       3 | u[0]*u[0]*u[0]                |         0 |
|       4 | d_dx(u[0],1)                  |         0 |
|       5 | d_dy(u[0],1)                  |         0 |
|       6 | d_dx(u[0]*u[0],1)             |         0 |
|       7 | d_dy(u[0]*u[0],1)             |         0 |
|       8 | d_dx(u[0]*u[0]*u[0],1)        |         0 |
|       9 | d_dy(u[0]*u[0]*u[0],1)        |         0 |
|      10 | d_dx(u[0],2)                  |         0 |
|      11 | d_dy(u[0],2)                  |         0 |
|      12 | dd_dxdy(u[0],(1,1))           |         0 |
|      13 | d_dx(u[0]*u[0],2)             |         0 |
|      14 | d_dy(u[0]*u[0],2)             |         0 |
|      15 | dd_dxdy(u[0]*u[0],(1,1))      |         0 |
[...]
\end{pyprint}
Of course just after initialization the coefficients of all these library terms are yet to be determined and are hence set to zero. The library supports some basic manipulations at this point, for more corresponding details see online tutorials. Now we can call the optimization routine to find the best sparse combination of the library functions. 
\begin{pyin}
    estimated_model.optimize_sigma(max_evals=10)
\end{pyin}

\begin{pyprint}
Generating library functions (this may take some time)...
100%|##########| 10/10 [08:13<00:00, 49.39s/trial, 
                        best loss: 3438.0777397909883]
Optimal threshold(s) found: {'h0': 0.8189787983433431}
New Sigma set to: 

|   Index | Term                   |   Value 0 |
|---------|------------------------|-----------|
|      10 | d_dx(u[0],2)           | -0.991169 |
|      11 | d_dy(u[0],2)           | -0.993026 |
|      16 | d_dx(u[0]*u[0]*u[0],2) |  0.983057 |
|      17 | d_dy(u[0]*u[0]*u[0],2) |  0.982041 |
|      31 | d_dx(u[0],4)           | -1.01657  |
|      32 | d_dy(u[0],4)           | -1.01647  |
|      35 | dd_dxdy(u[0],(2,2))    | -1.91144  |
    
\end{pyprint}
We find the results also reported in section~\ref{ssec:cahn_hilliard}, which agree very well with the original model. Note that the optimization routine itself has hyperparameters (such as the maximum number of evaluations) that will be passed to the tree-structured Parzen estimator of Hyperopt.
\newpage

\section{Detailed results for coefficients}\label{apx:results}
In this section we give unabbreviated numerical results for all computations performed throughout section~\ref{sec:results}, this includes all terms in our ansatz library, their estimated coefficients and the true model coefficients. All library terms that have all zero coefficients are listed as 'discarded terms', while terms with any non-zero coefficient for a field variable are listed in a table similar to the numerical output from the tutorial in section~\ref{apx:che_tutorial}, and compared to a table with the true model coefficients.
%\newpage
\begin{table}[ht]
    \begin{minipage}[t]{0.48\textwidth}
        \centering
        Estimated model \\
        \vspace{5pt}
\begin{tabular}{rcrr}
\hline
   Index & Term                                                                       &    $\partial_t u_1$ &    $\partial_t u_2$ \\
\hline
       1 & $u_{1}^{1}$                                                                &  0.983454  &  0         \\
       2 & $u_{2}^{1}$                                                                &  0         &  0.983574  \\
       6 & $u_{1}^{3}$                                                                & -0.983372  & -0.999644  \\
       7 & $u_{1}^{2}\,u_{2}^{1}$                                                     &  0.999632  & -0.98347   \\
       8 & $u_{1}^{1}\,u_{2}^{2}$                                                     & -0.983362  & -0.99963   \\
       9 & $u_{2}^{3}$                                                                &  0.999646  & -0.98349   \\
      12 & $\partial_y^{2}\left(u_{1}^{1}\right)$                                     &  0.0982818 &  0         \\
      13 & $\partial_y^{2}\left(u_{2}^{1}\right)$                                     &  0         &  0.0986993 \\
      18 & $\partial_x^{2}\left(u_{1}^{1}\right)$                                     &  0.0986863 &  0         \\
      19 & $\partial_x^{2}\left(u_{2}^{1}\right)$                                     &  0         &  0.0982933 \\
\hline
\end{tabular}

    \end{minipage}
    \hfill
    \begin{minipage}[t]{0.48\textwidth}
        \centering
        True model\\
        \vspace{5pt}
        \begin{tabular}{rcrr}
\hline
   Index & Term                                                                       &    $\partial_t u_1$ &    $\partial_t u_2$ \\
\hline
       1 & $u_{1}^{1}$                                                                &  1.0  &  0         \\
       2 & $u_{2}^{1}$                                                                &  0         &  1.0  \\
       6 & $u_{1}^{3}$                                                                & -1.0  & -1.0  \\
       7 & $u_{1}^{2}\,u_{2}^{1}$                                                     &  1.0  & -1.0   \\
       8 & $u_{1}^{1}\,u_{2}^{2}$                                                     & -1.0  & -1.0   \\
       9 & $u_{2}^{3}$                                                                &  1.0  & -1.0   \\
      12 & $\partial_y^{2}\left(u_{1}^{1}\right)$                                     &  0.1 &  0         \\
      13 & $\partial_y^{2}\left(u_{2}^{1}\right)$                                     &  0         &  0.1 \\
      18 & $\partial_x^{2}\left(u_{1}^{1}\right)$                                     &  0.1 &  0         \\
      19 & $\partial_x^{2}\left(u_{2}^{1}\right)$                                     &  0         &  0.1 \\
\hline
    \end{tabular}

    \end{minipage}

    \vfill
    \begin{minipage}[t]{0.9\textwidth}
        \vspace{10pt}
        \hrule
        \vspace{5pt}
        100 Discarded Terms:\\
        $1.0$, $u_{1}^{2}$, $u_{1}^{1}\,u_{2}^{1}$, $u_{2}^{2}$, $\partial_y^{1}\left(u_{1}^{1}\right)$, $\partial_y^{1}\left(u_{2}^{1}\right)$, $\partial_x^{1}\left(u_{1}^{1}\right)$, $\partial_x^{1}\left(u_{2}^{1}\right)$, $\partial_x^{1}\partial_y^{1}\left(u_{1}^{1}\right)$, $\partial_x^{1}\partial_y^{1}\left(u_{2}^{1}\right)$, $u_{1}^{1}\,\partial_y^{1}\left(u_{1}^{1}\right)$, $u_{2}^{1}\,\partial_y^{1}\left(u_{1}^{1}\right)$, $u_{1}^{2}\,\partial_y^{1}\left(u_{1}^{1}\right)$, $u_{1}^{1}\,u_{2}^{1}\,\partial_y^{1}\left(u_{1}^{1}\right)$, $u_{2}^{2}\,\partial_y^{1}\left(u_{1}^{1}\right)$, $u_{1}^{3}\,\partial_y^{1}\left(u_{1}^{1}\right)$, $u_{1}^{2}\,u_{2}^{1}\,\partial_y^{1}\left(u_{1}^{1}\right)$, $u_{1}^{1}\,u_{2}^{2}\,\partial_y^{1}\left(u_{1}^{1}\right)$, $u_{2}^{3}\,\partial_y^{1}\left(u_{1}^{1}\right)$, $u_{1}^{1}\,\partial_y^{1}\left(u_{2}^{1}\right)$, $u_{2}^{1}\,\partial_y^{1}\left(u_{2}^{1}\right)$, $u_{1}^{2}\,\partial_y^{1}\left(u_{2}^{1}\right)$, $u_{1}^{1}\,u_{2}^{1}\,\partial_y^{1}\left(u_{2}^{1}\right)$, $u_{2}^{2}\,\partial_y^{1}\left(u_{2}^{1}\right)$, $u_{1}^{3}\,\partial_y^{1}\left(u_{2}^{1}\right)$, $u_{1}^{2}\,u_{2}^{1}\,\partial_y^{1}\left(u_{2}^{1}\right)$, $u_{1}^{1}\,u_{2}^{2}\,\partial_y^{1}\left(u_{2}^{1}\right)$, $u_{2}^{3}\,\partial_y^{1}\left(u_{2}^{1}\right)$, $u_{1}^{1}\,\partial_y^{2}\left(u_{1}^{1}\right)$, $u_{2}^{1}\,\partial_y^{2}\left(u_{1}^{1}\right)$, $u_{1}^{2}\,\partial_y^{2}\left(u_{1}^{1}\right)$, $u_{1}^{1}\,u_{2}^{1}\,\partial_y^{2}\left(u_{1}^{1}\right)$, $u_{2}^{2}\,\partial_y^{2}\left(u_{1}^{1}\right)$, $u_{1}^{3}\,\partial_y^{2}\left(u_{1}^{1}\right)$, $u_{1}^{2}\,u_{2}^{1}\,\partial_y^{2}\left(u_{1}^{1}\right)$, $u_{1}^{1}\,u_{2}^{2}\,\partial_y^{2}\left(u_{1}^{1}\right)$, $u_{2}^{3}\,\partial_y^{2}\left(u_{1}^{1}\right)$, $u_{1}^{1}\,\partial_y^{2}\left(u_{2}^{1}\right)$, $u_{2}^{1}\,\partial_y^{2}\left(u_{2}^{1}\right)$, $u_{1}^{2}\,\partial_y^{2}\left(u_{2}^{1}\right)$, $u_{1}^{1}\,u_{2}^{1}\,\partial_y^{2}\left(u_{2}^{1}\right)$, $u_{2}^{2}\,\partial_y^{2}\left(u_{2}^{1}\right)$, $u_{1}^{3}\,\partial_y^{2}\left(u_{2}^{1}\right)$, $u_{1}^{2}\,u_{2}^{1}\,\partial_y^{2}\left(u_{2}^{1}\right)$, $u_{1}^{1}\,u_{2}^{2}\,\partial_y^{2}\left(u_{2}^{1}\right)$, $u_{2}^{3}\,\partial_y^{2}\left(u_{2}^{1}\right)$, $u_{1}^{1}\,\partial_x^{1}\left(u_{1}^{1}\right)$, $u_{2}^{1}\,\partial_x^{1}\left(u_{1}^{1}\right)$, $u_{1}^{2}\,\partial_x^{1}\left(u_{1}^{1}\right)$, $u_{1}^{1}\,u_{2}^{1}\,\partial_x^{1}\left(u_{1}^{1}\right)$, $u_{2}^{2}\,\partial_x^{1}\left(u_{1}^{1}\right)$, $u_{1}^{3}\,\partial_x^{1}\left(u_{1}^{1}\right)$, $u_{1}^{2}\,u_{2}^{1}\,\partial_x^{1}\left(u_{1}^{1}\right)$, $u_{1}^{1}\,u_{2}^{2}\,\partial_x^{1}\left(u_{1}^{1}\right)$, $u_{2}^{3}\,\partial_x^{1}\left(u_{1}^{1}\right)$, $u_{1}^{1}\,\partial_x^{1}\left(u_{2}^{1}\right)$, $u_{2}^{1}\,\partial_x^{1}\left(u_{2}^{1}\right)$, $u_{1}^{2}\,\partial_x^{1}\left(u_{2}^{1}\right)$, $u_{1}^{1}\,u_{2}^{1}\,\partial_x^{1}\left(u_{2}^{1}\right)$, $u_{2}^{2}\,\partial_x^{1}\left(u_{2}^{1}\right)$, $u_{1}^{3}\,\partial_x^{1}\left(u_{2}^{1}\right)$, $u_{1}^{2}\,u_{2}^{1}\,\partial_x^{1}\left(u_{2}^{1}\right)$, $u_{1}^{1}\,u_{2}^{2}\,\partial_x^{1}\left(u_{2}^{1}\right)$, $u_{2}^{3}\,\partial_x^{1}\left(u_{2}^{1}\right)$, $u_{1}^{1}\,\partial_x^{1}\partial_y^{1}\left(u_{1}^{1}\right)$, $u_{2}^{1}\,\partial_x^{1}\partial_y^{1}\left(u_{1}^{1}\right)$, $u_{1}^{2}\,\partial_x^{1}\partial_y^{1}\left(u_{1}^{1}\right)$, $u_{1}^{1}\,u_{2}^{1}\,\partial_x^{1}\partial_y^{1}\left(u_{1}^{1}\right)$, $u_{2}^{2}\,\partial_x^{1}\partial_y^{1}\left(u_{1}^{1}\right)$, $u_{1}^{3}\,\partial_x^{1}\partial_y^{1}\left(u_{1}^{1}\right)$, $u_{1}^{2}\,u_{2}^{1}\,\partial_x^{1}\partial_y^{1}\left(u_{1}^{1}\right)$, $u_{1}^{1}\,u_{2}^{2}\,\partial_x^{1}\partial_y^{1}\left(u_{1}^{1}\right)$, $u_{2}^{3}\,\partial_x^{1}\partial_y^{1}\left(u_{1}^{1}\right)$, $u_{1}^{1}\,\partial_x^{1}\partial_y^{1}\left(u_{2}^{1}\right)$, $u_{2}^{1}\,\partial_x^{1}\partial_y^{1}\left(u_{2}^{1}\right)$, $u_{1}^{2}\,\partial_x^{1}\partial_y^{1}\left(u_{2}^{1}\right)$, $u_{1}^{1}\,u_{2}^{1}\,\partial_x^{1}\partial_y^{1}\left(u_{2}^{1}\right)$, $u_{2}^{2}\,\partial_x^{1}\partial_y^{1}\left(u_{2}^{1}\right)$, $u_{1}^{3}\,\partial_x^{1}\partial_y^{1}\left(u_{2}^{1}\right)$, $u_{1}^{2}\,u_{2}^{1}\,\partial_x^{1}\partial_y^{1}\left(u_{2}^{1}\right)$, $u_{1}^{1}\,u_{2}^{2}\,\partial_x^{1}\partial_y^{1}\left(u_{2}^{1}\right)$, $u_{2}^{3}\,\partial_x^{1}\partial_y^{1}\left(u_{2}^{1}\right)$, $u_{1}^{1}\,\partial_x^{2}\left(u_{1}^{1}\right)$, $u_{2}^{1}\,\partial_x^{2}\left(u_{1}^{1}\right)$, $u_{1}^{2}\,\partial_x^{2}\left(u_{1}^{1}\right)$, $u_{1}^{1}\,u_{2}^{1}\,\partial_x^{2}\left(u_{1}^{1}\right)$, $u_{2}^{2}\,\partial_x^{2}\left(u_{1}^{1}\right)$, $u_{1}^{3}\,\partial_x^{2}\left(u_{1}^{1}\right)$, $u_{1}^{2}\,u_{2}^{1}\,\partial_x^{2}\left(u_{1}^{1}\right)$, $u_{1}^{1}\,u_{2}^{2}\,\partial_x^{2}\left(u_{1}^{1}\right)$, $u_{2}^{3}\,\partial_x^{2}\left(u_{1}^{1}\right)$, $u_{1}^{1}\,\partial_x^{2}\left(u_{2}^{1}\right)$, $u_{2}^{1}\,\partial_x^{2}\left(u_{2}^{1}\right)$, $u_{1}^{2}\,\partial_x^{2}\left(u_{2}^{1}\right)$, $u_{1}^{1}\,u_{2}^{1}\,\partial_x^{2}\left(u_{2}^{1}\right)$, $u_{2}^{2}\,\partial_x^{2}\left(u_{2}^{1}\right)$, $u_{1}^{3}\,\partial_x^{2}\left(u_{2}^{1}\right)$, $u_{1}^{2}\,u_{2}^{1}\,\partial_x^{2}\left(u_{2}^{1}\right)$, $u_{1}^{1}\,u_{2}^{2}\,\partial_x^{2}\left(u_{2}^{1}\right)$, $u_{2}^{3}\,\partial_x^{2}\left(u_{2}^{1}\right)$ \\
        \vspace{5pt}
        \hrule

    \end{minipage}

    \caption{Full results for estimating the reaction diffusion equation (\ref{eqn:rd_1}) from data shown in figure~\ref{fig:rd_and_estimate_selection} from section~\ref{sssec:baseline}. Identical results for both our method (found threshold: $h=0.0811991935628$) and \emph{PySINDy} (threshold: $h = 0.08$)}
    \label{tab:apx_rd}
\end{table}

\begin{table}[ht]
    \begin{minipage}[t]{0.48\textwidth}
        \centering
        Estimated model \\
        \vspace{5pt}
        \begin{tabular}{rcr}
        \hline
        Index & Term                                                 &   $\partial_t u_1$ \\
        \hline
       1 & $u_{0}^{1}$                                          &  0.98642  \\
       3 & $u_{0}^{3}$                                          & -0.986013 \\
      10 & $\partial_x^{2}\left(u_{0}^{1}\right)$               &  0.987685 \\
      11 & $\partial_y^{2}\left(u_{0}^{1}\right)$               &  0.987623 \\
      \hline
    \end{tabular}

    \end{minipage}
    \hfill
    \begin{minipage}[t]{0.48\textwidth}
        \centering
        True model \\
        \vspace{5pt}
        \begin{tabular}{rcr}
        \hline
        Index & Term                                                 &   $\partial_t u_1$ \\
        \hline
       1 & $u_{0}^{1}$                                          &  1  \\
       3 & $u_{0}^{3}$                                          & -1 \\
      10 & $\partial_x^{2}\left(u_{0}^{1}\right)$               &  1 \\
      11 & $\partial_y^{2}\left(u_{0}^{1}\right)$               &  1 \\
      \hline
    \end{tabular}

    \end{minipage}
    
    \vfill
    \begin{minipage}[t]{0.9\textwidth}
        \vspace{10pt}
        \hrule
        \vspace{5pt}
        42 Discarded Terms:\\
        $1.0$, $u_{0}^{1}$, $\partial_x^{1}\left(u_{0}^{1}\right)$, $\partial_y^{1}\left(u_{0}^{1}\right)$, $\partial_x^{1}\left(u_{0}^{2}\right)$, $\partial_y^{1}\left(u_{0}^{2}\right)$, $\partial_x^{1}\left(u_{0}^{3}\right)$, $\partial_y^{1}\left(u_{0}^{3}\right)$, $\partial_x^{1}\partial_y^{1}\left(u_{0}^{1}\right)$, $\partial_x^{2}\left(u_{0}^{2}\right)$, $\partial_y^{2}\left(u_{0}^{2}\right)$, $\partial_x^{1}\partial_y^{1}\left(u_{0}^{2}\right)$, $\partial_x^{2}\left(u_{0}^{3}\right)$, $\partial_y^{2}\left(u_{0}^{3}\right)$, $\partial_x^{1}\partial_y^{1}\left(u_{0}^{3}\right)$, $\partial_x^{3}\left(u_{0}^{1}\right)$, $\partial_y^{3}\left(u_{0}^{1}\right)$, $\partial_x^{1}\partial_y^{2}\left(u_{0}^{1}\right)$, $\partial_x^{2}\partial_y^{1}\left(u_{0}^{1}\right)$, $\partial_x^{3}\left(u_{0}^{2}\right)$, $\partial_y^{3}\left(u_{0}^{2}\right)$, $\partial_x^{1}\partial_y^{2}\left(u_{0}^{2}\right)$, $\partial_x^{2}\partial_y^{1}\left(u_{0}^{2}\right)$, $\partial_x^{3}\left(u_{0}^{3}\right)$, $\partial_y^{3}\left(u_{0}^{3}\right)$, $\partial_x^{1}\partial_y^{2}\left(u_{0}^{3}\right)$, $\partial_x^{2}\partial_y^{1}\left(u_{0}^{3}\right)$, $\partial_x^{4}\left(u_{0}^{1}\right)$, $\partial_y^{4}\left(u_{0}^{1}\right)$, $\partial_x^{1}\partial_y^{3}\left(u_{0}^{1}\right)$, $\partial_x^{3}\partial_y^{1}\left(u_{0}^{1}\right)$, $\partial_x^{2}\partial_y^{2}\left(u_{0}^{1}\right)$, $\partial_x^{4}\left(u_{0}^{2}\right)$, $\partial_y^{4}\left(u_{0}^{2}\right)$, $\partial_x^{1}\partial_y^{3}\left(u_{0}^{2}\right)$, $\partial_x^{3}\partial_y^{1}\left(u_{0}^{2}\right)$, $\partial_x^{2}\partial_y^{2}\left(u_{0}^{2}\right)$, $\partial_x^{4}\left(u_{0}^{3}\right)$, $\partial_y^{4}\left(u_{0}^{3}\right)$, $\partial_x^{1}\partial_y^{3}\left(u_{0}^{3}\right)$, $\partial_x^{3}\partial_y^{1}\left(u_{0}^{3}\right)$, $\partial_x^{2}\partial_y^{2}\left(u_{0}^{3}\right)$
\\
        \vspace{5pt}
        \hrule

    \end{minipage}

    \caption{Full results for estimating the Allen-Cahn equation (\ref{eqn:ACE}) from data shown in figure~\ref{fig:ace_selection} from section~\ref{ssec:allen_cahn}. Found hyperparameter (threshold): $h_1 = 0.3300998462576791$.}
    \label{tab:apx_ace}
\end{table}

\begin{table}[ht]
    \begin{minipage}[t]{0.48\textwidth}
        \centering
        Estimated model \\
        \vspace{5pt}
        \begin{tabular}{rcr}
        \hline
        Index & Term                                                 &   $\partial_t u_1$ \\
        \hline
      10 & $\partial_x^{2}\left(u_{1}^{1}\right)$               & -0.986104 \\
      11 & $\partial_y^{2}\left(u_{1}^{1}\right)$               & -0.989164 \\
      16 & $\partial_x^{2}\left(u_{1}^{3}\right)$               &  0.977613 \\
      17 & $\partial_y^{2}\left(u_{1}^{3}\right)$               &  0.985195 \\
      31 & $\partial_x^{4}\left(u_{1}^{1}\right)$               & -1.02871  \\
      32 & $\partial_y^{4}\left(u_{1}^{1}\right)$               & -1.0264   \\
      35 & $\partial_x^{2}\partial_y^{2}\left(u_{1}^{1}\right)$ & -1.91485  \\
      \hline
    \end{tabular}

    \end{minipage}
    \hfill
    \begin{minipage}[t]{0.48\textwidth}
        \centering
        True model \\
        \vspace{5pt}
        \begin{tabular}{rcr}
        \hline
        Index & Term                                                 &   $\partial_t u_1$ \\
        \hline
      10 & $\partial_x^{2}\left(u_{1}^{1}\right)$               & -1.0 \\
      11 & $\partial_y^{2}\left(u_{1}^{1}\right)$               & -1.0 \\
      16 & $\partial_x^{2}\left(u_{1}^{3}\right)$               &  1.0 \\
      17 & $\partial_y^{2}\left(u_{1}^{3}\right)$               &  1.0 \\
      31 & $\partial_x^{4}\left(u_{1}^{1}\right)$               & -1.0  \\
      32 & $\partial_y^{4}\left(u_{1}^{1}\right)$               & -1.0   \\
      35 & $\partial_x^{2}\partial_y^{2}\left(u_{1}^{1}\right)$ & -2.0  \\
      \hline
    \end{tabular}

    \end{minipage}
    
    \vfill
    \begin{minipage}[t]{0.9\textwidth}
        \vspace{10pt}
        \hrule
        \vspace{5pt}
        39 Discarded Terms:\\
        $1.0$, $u_{1}^{1}$, $u_{2}^{1}$, $u_{1}^{2}$, $u_{1}^{1}\,u_{2}^{1}$, $u_{2}^{2}$, $u_{1}^{3}$, $u_{1}^{1}\,u_{2}^{2}$, $u_{1}^{2}\,u_{2}^{1}$, $u_{2}^{3}$, $\partial_x^{1}\left(u_{2}^{1}\right)$, $\partial_y^{1}\left(u_{2}^{1}\right)$, $\partial_x^{1}\left(u_{1}^{2}\right)$, $\partial_y^{1}\left(u_{1}^{2}\right)$, $\partial_x^{1}\left(u_{2}^{2}\right)$, $\partial_y^{1}\left(u_{2}^{2}\right)$, $\partial_x^{1}\left(u_{1}^{3}\right)$, $\partial_y^{1}\left(u_{1}^{3}\right)$, $\partial_x^{1}\left(u_{1}^{1}\,u_{2}^{2}\right)$, $\partial_y^{1}\left(u_{1}^{1}\,u_{2}^{2}\right)$, $\partial_x^{1}\left(u_{1}^{2}\,u_{2}^{1}\right)$, $\partial_y^{1}\left(u_{1}^{2}\,u_{2}^{1}\right)$, $\partial_x^{1}\left(u_{2}^{3}\right)$, $\partial_y^{1}\left(u_{2}^{3}\right)$, $\partial_x^{2}\left(u_{1}^{1}\right)$, $\partial_y^{2}\left(u_{1}^{1}\right)$, $\partial_x^{1}\partial_y^{1}\left(u_{1}^{1}\right)$, $\partial_x^{1}\partial_y^{1}\left(u_{2}^{1}\right)$, $\partial_x^{2}\left(u_{1}^{2}\right)$, $\partial_x^{1}\partial_y^{1}\left(u_{1}^{2}\right)$, $\partial_x^{2}\left(u_{1}^{1}\,u_{2}^{1}\right)$, $\partial_y^{2}\left(u_{1}^{1}\,u_{2}^{1}\right)$, $\partial_x^{1}\partial_y^{1}\left(u_{1}^{1}\,u_{2}^{1}\right)$, $\partial_x^{2}\left(u_{2}^{2}\right)$, $\partial_y^{2}\left(u_{2}^{2}\right)$, $\partial_x^{1}\partial_y^{1}\left(u_{2}^{2}\right)$, $\partial_x^{2}\left(u_{1}^{3}\right)$, $\partial_y^{2}\left(u_{1}^{3}\right)$, $\partial_x^{1}\partial_y^{1}\left(u_{1}^{3}\right)$\\
        \vspace{5pt}
        \hrule

    \end{minipage}

    \caption{Full results for estimating the Cahn-Hilliard equation (\ref{eqn:CHE}) from data shown in figure~\ref{fig:che_selection} from section~\ref{ssec:cahn_hilliard}. Found hyperparameter (threshold): $h_1 =  0.33003545771798815$.}
    \label{tab:apx_che}
\end{table}

\begin{table}[ht]
    \begin{minipage}[t]{0.48\textwidth}
        \centering
        Estimated model \\
        \vspace{5pt}
        \begin{tabular}{rcrr}
        \hline
        Index & Term                                                       &   $\partial_t u_1$ &    $\partial_t u_2$ \\
        \hline
       1 & $u_{1}^{1}$                                                     &  0.485341 &  0.0664892 \\
       2 & $u_{2}^{1}$                                                     & -0.493396 & -0.0664834 \\
       6 & $u_{1}^{3}$                                                     & -0.965856 &  0         \\
       28 & $\partial_x^{2}\left(u_{1}^{1}\right)$                          &  0.242025 &  0         \\
      29 & $\partial_y^{2}\left(u_{1}^{1}\right)$                          &  0.242146 &  0         \\
      31 & $\partial_x^{2}\left(u_{2}^{1}\right)$                          &  0        &  0.0167489 \\
      32 & $\partial_y^{2}\left(u_{2}^{1}\right)$                          &  0        &  0.0167483 \\
      \hline
    \end{tabular}

    \end{minipage}
    \hfill
    \begin{minipage}[t]{0.48\textwidth}
        \centering
        True model\\
        \vspace{5pt}
        \begin{tabular}{rcrr}
        \hline
        Index & Term                                                       &   $\partial_t u_1$ &    $\partial_t u_2$ \\
        \hline
       1 & $u_{1}^{1}$                                                     &  0.5 &  0.066 \\
       2 & $u_{2}^{1}$                                                     & -0.5 & -0.066 \\
       6 & $u_{1}^{3}$                                                     & -1.0 &  0         \\
       28 & $\partial_x^{2}\left(u_{1}^{1}\right)$                          &  0.5 &  0         \\
      29 & $\partial_y^{2}\left(u_{1}^{1}\right)$                          &  0.5 &  0         \\
      31 & $\partial_x^{2}\left(u_{2}^{1}\right)$                          &  0        &  0.0166 \\
      32 & $\partial_y^{2}\left(u_{2}^{1}\right)$                          &  0        &  0.0166 \\
      \hline
    \end{tabular}

    \end{minipage}

    \vfill
    \begin{minipage}[t]{0.9\textwidth}
        \vspace{10pt}
        \hrule
        \vspace{5pt}
        48 Discarded Terms:\\
        $1.0$, $u_{1}^{2}$, $u_{1}^{1}\,u_{2}^{1}$, $u_{2}^{2}$, $u_{1}^{1}\,u_{2}^{2}$, $u_{1}^{2}\,u_{2}^{1}$, $u_{2}^{3}$, $\partial_x^{1}\left(u_{1}^{1}\right)$, $\partial_y^{1}\left(u_{1}^{1}\right)$, $\partial_x^{1}\left(u_{2}^{1}\right)$, $\partial_y^{1}\left(u_{2}^{1}\right)$, $\partial_x^{1}\left(u_{1}^{2}\right)$, $\partial_y^{1}\left(u_{1}^{2}\right)$, $\partial_x^{1}\left(u_{1}^{1}\,u_{2}^{1}\right)$, $\partial_y^{1}\left(u_{1}^{1}\,u_{2}^{1}\right)$, $\partial_x^{1}\left(u_{2}^{2}\right)$, $\partial_y^{1}\left(u_{2}^{2}\right)$, $\partial_x^{1}\left(u_{1}^{3}\right)$, $\partial_y^{1}\left(u_{1}^{3}\right)$, $\partial_x^{1}\left(u_{1}^{1}\,u_{2}^{2}\right)$, $\partial_y^{1}\left(u_{1}^{1}\,u_{2}^{2}\right)$, $\partial_x^{1}\left(u_{1}^{2}\,u_{2}^{1}\right)$, $\partial_y^{1}\left(u_{1}^{2}\,u_{2}^{1}\right)$, $\partial_x^{1}\left(u_{2}^{3}\right)$, $\partial_y^{1}\left(u_{2}^{3}\right)$, $\partial_x^{1}\partial_y^{1}\left(u_{1}^{1}\right)$, $\partial_x^{1}\partial_y^{1}\left(u_{2}^{1}\right)$, $\partial_x^{2}\left(u_{1}^{2}\right)$, $\partial_y^{2}\left(u_{1}^{2}\right)$, $\partial_x^{1}\partial_y^{1}\left(u_{1}^{2}\right)$, $\partial_x^{2}\left(u_{1}^{1}\,u_{2}^{1}\right)$, $\partial_y^{2}\left(u_{1}^{1}\,u_{2}^{1}\right)$, $\partial_x^{1}\partial_y^{1}\left(u_{1}^{1}\,u_{2}^{1}\right)$, $\partial_x^{2}\left(u_{2}^{2}\right)$, $\partial_y^{2}\left(u_{2}^{2}\right)$, $\partial_x^{1}\partial_y^{1}\left(u_{2}^{2}\right)$, $\partial_x^{2}\left(u_{1}^{3}\right)$, $\partial_y^{2}\left(u_{1}^{3}\right)$, $\partial_x^{1}\partial_y^{1}\left(u_{1}^{3}\right)$, $\partial_x^{2}\left(u_{1}^{1}\,u_{2}^{2}\right)$, $\partial_y^{2}\left(u_{1}^{1}\,u_{2}^{2}\right)$, $\partial_x^{1}\partial_y^{1}\left(u_{1}^{1}\,u_{2}^{2}\right)$, $\partial_x^{2}\left(u_{1}^{2}\,u_{2}^{1}\right)$, $\partial_y^{2}\left(u_{1}^{2}\,u_{2}^{1}\right)$, $\partial_x^{1}\partial_y^{1}\left(u_{1}^{2}\,u_{2}^{1}\right)$, $\partial_x^{2}\left(u_{2}^{3}\right)$, $\partial_y^{2}\left(u_{2}^{3}\right)$, $\partial_x^{1}\partial_y^{1}\left(u_{2}^{3}\right)$ \\
        \vspace{5pt}
        \hrule

    \end{minipage}

    \caption{Full results for estimating the FitzHugh-Nagumo equation (\ref{eqn:FHN_u}) from data shown in figure~\ref{fig:fhn_estimate_selection} from section~\ref{ssec:fhn}. Found hyperparameter (thresholds): $h_1= 0.06642802839932697$ for all terms for $\partial_t u_1$ and $h_2 = 0.013702806950437454$ for all terms for $\partial_t u_2$.}
    \label{tab:apx_fhn}
\end{table}

\begin{table}[ht]
    \begin{minipage}[t]{0.48\textwidth}
        \centering
        Estimated model\\
        \vspace{5pt}
        \begin{tabular}{rcrr}
            \hline
            Index & Term &   $\partial_t u_1$ &   $\partial_t u_2$ \\
            \hline
            1 & $u_{1}^{1}$                                                     &  0.999621 &  0        \\
            2 & $u_{2}^{1}$                                                     &  0        &  0.999654 \\
            6 & $u_{1}^{3}$                                                     & -0.999402 & 3.99925   \\
            7 & $u_{1}^{2}\,u_{1}^{1}$                                          & -3.99919   & -0.999489 \\
            8 & $u_{1}^{1}\,u_{1}^{2}$                                          & -0.999413 & 3.99925  \\
            9 & $u_{2}^{3}$                                                     & -3.99918   & -0.99947 \\
            28 & $\partial_x^{2}\left(u_{1}^{1}\right)$                          &  0.977855 &  1.99941        \\
            29 & $\partial_y^{2}\left(u_{1}^{1}\right)$                          &  0.983671 &  1.9994        \\
            31 & $\partial_x^{2}\left(u_{2}^{1}\right)$                          &  -1.99934        &  0.974819 \\
            32 & $\partial_y^{2}\left(u_{2}^{1}\right)$                          &  -1.99935        &  0.982875 \\
            \hline
        \end{tabular}
    \end{minipage}
    \hfill
    \begin{minipage}[t]{0.48\textwidth}
        \centering
        True model\\
        \vspace{5pt}
        \begin{tabular}{rcrr}
            \hline
            Index & Term &   $\partial_t u_1$ &   $\partial_t u_2$ \\
            \hline
            1 & $u_{1}^{1}$                                                     &  1.0 &  0        \\
            2 & $u_{2}^{1}$                                                     &  0        &  1.0 \\
            6 & $u_{1}^{3}$                                                     & -1.0 & 4.0   \\
            7 & $u_{1}^{2}\,u_{1}^{1}$                                          & -4.0   & -1.0 \\
            8 & $u_{1}^{1}\,u_{1}^{2}$                                          & -1.0 & 4.0  \\
            9 & $u_{2}^{3}$                                                     & -4.0   & -1.0 \\
            28 & $\partial_x^{2}\left(u_{1}^{1}\right)$                          &  1.0 &  2.0        \\
            29 & $\partial_y^{2}\left(u_{1}^{1}\right)$                          &  1.0 &  2.0        \\
            31 & $\partial_x^{2}\left(u_{2}^{1}\right)$                          &  -2.0        &  1.0 \\
            32 & $\partial_y^{2}\left(u_{2}^{1}\right)$                          &  -2.0        &  1.0 \\
            \hline
        \end{tabular}
    \end{minipage}
    \vfill
    \begin{minipage}[t]{0.9\textwidth}
        \vspace{10pt}
        \hrule
        \vspace{5pt}
        45 Discarded Terms:\\
        $1.0$, $u_{1}^{2}$, $u_{1}^{1}\,u_{2}^{1}$, $u_{2}^{2}$, $\partial_x^{1}\left(u_{1}^{1}\right)$, $\partial_y^{1}\left(u_{1}^{1}\right)$, $\partial_x^{1}\left(u_{2}^{1}\right)$, $\partial_y^{1}\left(u_{2}^{1}\right)$, $\partial_x^{1}\left(u_{1}^{2}\right)$, $\partial_y^{1}\left(u_{1}^{2}\right)$, $\partial_x^{1}\left(u_{1}^{1}\,u_{2}^{1}\right)$, $\partial_y^{1}\left(u_{1}^{1}\,u_{2}^{1}\right)$, $\partial_x^{1}\left(u_{2}^{2}\right)$, $\partial_y^{1}\left(u_{2}^{2}\right)$, $\partial_x^{1}\left(u_{1}^{3}\right)$, $\partial_y^{1}\left(u_{1}^{3}\right)$, $\partial_x^{1}\left(u_{1}^{1}\,u_{2}^{2}\right)$, $\partial_y^{1}\left(u_{1}^{1}\,u_{2}^{2}\right)$, $\partial_x^{1}\left(u_{1}^{2}\,u_{2}^{1}\right)$, $\partial_y^{1}\left(u_{1}^{2}\,u_{2}^{1}\right)$, $\partial_x^{1}\left(u_{2}^{3}\right)$, $\partial_y^{1}\left(u_{2}^{3}\right)$, $\partial_x^{1}\partial_y^{1}\left(u_{1}^{1}\right)$, $\partial_x^{1}\partial_y^{1}\left(u_{2}^{1}\right)$, $\partial_x^{2}\left(u_{1}^{2}\right)$, $\partial_y^{2}\left(u_{1}^{2}\right)$, $\partial_x^{1}\partial_y^{1}\left(u_{1}^{2}\right)$, $\partial_x^{2}\left(u_{1}^{1}\,u_{2}^{1}\right)$, $\partial_y^{2}\left(u_{1}^{1}\,u_{2}^{1}\right)$, $\partial_x^{1}\partial_y^{1}\left(u_{1}^{1}\,u_{2}^{1}\right)$, $\partial_x^{2}\left(u_{2}^{2}\right)$, $\partial_y^{2}\left(u_{2}^{2}\right)$, $\partial_x^{1}\partial_y^{1}\left(u_{2}^{2}\right)$, $\partial_x^{2}\left(u_{1}^{3}\right)$, $\partial_y^{2}\left(u_{1}^{3}\right)$, $\partial_x^{1}\partial_y^{1}\left(u_{1}^{3}\right)$, $\partial_x^{2}\left(u_{1}^{1}\,u_{2}^{2}\right)$, $\partial_y^{2}\left(u_{1}^{1}\,u_{2}^{2}\right)$, $\partial_x^{1}\partial_y^{1}\left(u_{1}^{1}\,u_{2}^{2}\right)$, $\partial_x^{2}\left(u_{1}^{2}\,u_{2}^{1}\right)$, $\partial_y^{2}\left(u_{1}^{2}\,u_{2}^{1}\right)$, $\partial_x^{1}\partial_y^{1}\left(u_{1}^{2}\,u_{2}^{1}\right)$, $\partial_x^{2}\left(u_{2}^{3}\right)$, $\partial_y^{2}\left(u_{2}^{3}\right)$, $\partial_x^{1}\partial_y^{1}\left(u_{2}^{3}\right)$\\
        \vspace{5pt}
        \hrule

    \end{minipage}

    \caption{Full results for estimating the reaction diffusion equations (\ref{eqn:rdc_1}) from data shown in figure~\ref{fig:rd_chaotic_1} from section~\ref{ssec:rd_chaotic}. Found hyperparameters (thresholds) are: $h_1 =  0.3066755436690512$ for all terms for $\partial_t u_1$ and $h_2 = 0.17758353931933746$ for all terms for $\partial_t u_2$.}
    \label{tab:apx_rd_chaotic_1}
\end{table}

\begin{table}[]
    \begin{minipage}[t]{0.48\textwidth}
        \centering
        Estimated model \\
        \vspace{5pt}
        \begin{tabular}{rcrr}
            \hline
            Index & Term &   $\partial_t u_1$ &   $\partial_t u_2$ \\
            \hline
       1 & $u_{1}^{1}$                                                     &  0.977683 &  0        \\
       2 & $u_{2}^{1}$                                                     &  0        &  0.979197 \\
       6 & $u_{1}^{3}$                                                     & -0.998367 &  3.99646  \\
       7 & $u_{2}^{1}\,u_{0}^{2}$                                          & -3.99625  & -0.980391 \\
       8 & $u_{2}^{2}\,u_{0}^{1}$                                          & -0.979247 &  3.99825  \\
       9 & $u_{2}^{3}$                                                     & -3.99615  & -1.00044  \\
      28 & $\partial_x^{2}\left(u_{1}^{1}\right)$                          &  0.978702 &  0        \\
      29 & $\partial_y^{2}\left(u_{1}^{1}\right)$                          &  0.961488 &  0        \\
      31 & $\partial_x^{2}\left(u_{2}^{1}\right)$                          &  0        &  0.979449 \\
      32 & $\partial_y^{2}\left(u_{2}^{1}\right)$                          &  0        &  0.96695  \\
      44 & $\partial_y^{2}\left(u_{1}^{3}\right)$                          & -0.146276 &  0        \\
      47 & $\partial_y^{2}\left(u_{2}^{1}\,u_{0}^{2}\right)$               &  0        & -0.122473 \\
      50 & $\partial_y^{2}\left(u_{2}^{2}\,u_{0}^{1}\right)$               & -0.119995 &  0        \\
      53 & $\partial_y^{2}\left(u_{2}^{3}\right)$                          &  0        & -0.154368 \\
      \hline
        \end{tabular}
    \end{minipage}
    \hfill
    \begin{minipage}[t]{0.48\textwidth}
        \centering
        True model \\
        \vspace{5pt}
        \begin{tabular}{rcrr}
            \hline
            Index & Term &   $\partial_t u_1$ &   $\partial_t u_2$ \\
            \hline
            1 & $u_{1}^{1}$                                                     &  1.0 &  0        \\
            2 & $u_{2}^{1}$                                                     &  0        &  1.0 \\
            6 & $u_{1}^{3}$                                                     & -1.0 &  4.0  \\
            7 & $u_{1}^{2}\,u_{1}^{1}$                                          & -4.0  & -1.0 \\
            8 & $u_{1}^{1}\,u_{1}^{2}$                                          & -1.0 &  4.0  \\
            9 & $u_{2}^{3}$                                                     &-4.0    & -1.0 \\
            28 & $\partial_x^{2}\left(u_{1}^{1}\right)$                          &  1.0 &  0        \\
            29 & $\partial_y^{2}\left(u_{1}^{1}\right)$                          &  1.0 &  0        \\
            31 & $\partial_x^{2}\left(u_{2}^{1}\right)$                          &  0        &  1.0 \\
            32 & $\partial_y^{2}\left(u_{2}^{1}\right)$                          &  0        &  1.0 \\
            \hline
        \end{tabular}
    \end{minipage}
    \vfill
    \begin{minipage}[t]{0.9\textwidth}
        \vspace{10pt}
        \hrule
        \vspace{5pt}
        41 Discarded Terms:\\
        $1.0$, $u_{1}^{2}$, $u_{1}^{1}\,u_{2}^{1}$, $u_{2}^{2}$, $\partial_x^{1}\left(u_{1}^{1}\right)$, $\partial_y^{1}\left(u_{1}^{1}\right)$, $\partial_x^{1}\left(u_{2}^{1}\right)$, $\partial_y^{1}\left(u_{2}^{1}\right)$, $\partial_x^{1}\left(u_{1}^{2}\right)$, $\partial_y^{1}\left(u_{1}^{2}\right)$, $\partial_x^{1}\left(u_{1}^{1}\,u_{2}^{1}\right)$, $\partial_y^{1}\left(u_{1}^{1}\,u_{2}^{1}\right)$, $\partial_x^{1}\left(u_{2}^{2}\right)$, $\partial_y^{1}\left(u_{2}^{2}\right)$, $\partial_x^{1}\left(u_{1}^{3}\right)$, $\partial_y^{1}\left(u_{1}^{3}\right)$, $\partial_x^{1}\left(u_{1}^{1}\,u_{2}^{2}\right)$, $\partial_y^{1}\left(u_{1}^{1}\,u_{2}^{2}\right)$, $\partial_x^{1}\left(u_{1}^{2}\,u_{2}^{1}\right)$, $\partial_y^{1}\left(u_{1}^{2}\,u_{2}^{1}\right)$, $\partial_x^{1}\left(u_{2}^{3}\right)$, $\partial_y^{1}\left(u_{2}^{3}\right)$, $\partial_x^{1}\partial_y^{1}\left(u_{1}^{1}\right)$, $\partial_x^{1}\partial_y^{1}\left(u_{2}^{1}\right)$, $\partial_x^{2}\left(u_{1}^{2}\right)$, $\partial_y^{2}\left(u_{1}^{2}\right)$, $\partial_x^{1}\partial_y^{1}\left(u_{1}^{2}\right)$, $\partial_x^{2}\left(u_{1}^{1}\,u_{2}^{1}\right)$, $\partial_y^{2}\left(u_{1}^{1}\,u_{2}^{1}\right)$, $\partial_x^{1}\partial_y^{1}\left(u_{1}^{1}\,u_{2}^{1}\right)$, $\partial_x^{2}\left(u_{2}^{2}\right)$, $\partial_y^{2}\left(u_{2}^{2}\right)$, $\partial_x^{1}\partial_y^{1}\left(u_{2}^{2}\right)$, $\partial_x^{2}\left(u_{1}^{3}\right)$, $\partial_x^{1}\partial_y^{1}\left(u_{1}^{3}\right)$, $\partial_x^{2}\left(u_{1}^{1}\,u_{2}^{2}\right)$, $\partial_x^{1}\partial_y^{1}\left(u_{1}^{1}\,u_{2}^{2}\right)$, $\partial_x^{2}\left(u_{1}^{2}\,u_{2}^{1}\right)$, $\partial_x^{1}\partial_y^{1}\left(u_{1}^{2}\,u_{2}^{1}\right)$, $\partial_x^{2}\left(u_{2}^{3}\right)$, $\partial_x^{1}\partial_y^{1}\left(u_{2}^{3}\right)$\\
        \vspace{5pt}
        \hrule

    \end{minipage}

    \caption{Full results for estimating reaction diffusion equations (\ref{eqn:rdc_1}) from data shown in figure~\ref{fig:rd_chaotic_2} from section~\ref{ssec:rd_chaotic} without additional thresholds. Found hyperparameters (thresholds): $h_1= 0.3066755436690512$ for all terms for $\partial_t u_1$ and $h_2 =  0.17758353931933746$ for all terms for $\partial_t u_2$.}
    \label{tab:apx_rd_chaotic_2a}
\end{table}

\begin{table}[]
    \begin{minipage}[t]{0.48\textwidth}
        \centering
        Estimated model \\
        \vspace{5pt}
        \begin{tabular}{rcrr}
            \hline
            Index & Term &   $\partial_t u_1$ &   $\partial_t u_2$ \\
            \hline
            1 & $u_{1}^{1}$                                                     &  0.956482 &  0        \\
            2 & $u_{2}^{1}$                                                     &  0        &  0.95713 \\
            6 & $u_{1}^{3}$                                                     & -0.959624 &  3.99263  \\
            7 & $u_{1}^{2}\,u_{1}^{1}$                                          & -3.99189  & -0.960558 \\
            8 & $u_{1}^{1}\,u_{1}^{2}$                                          & -0.959277 &  3.99266  \\
            9 & $u_{2}^{3}$                                                     &-3.9922    & -0.959894 \\
            28 & $\partial_x^{2}\left(u_{1}^{1}\right)$                          &  0.974305 &  0        \\
            29 & $\partial_y^{2}\left(u_{1}^{1}\right)$                          &  0.89647 &  0        \\
            31 & $\partial_x^{2}\left(u_{2}^{1}\right)$                          &  0        &  0.974735 \\
            32 & $\partial_y^{2}\left(u_{2}^{1}\right)$                          &  0        &  0.898804 \\
            \hline
        \end{tabular}
    \end{minipage}
    \hfill
    \begin{minipage}[t]{0.48\textwidth}
        \centering
        True model \\
        \vspace{5pt}
        \begin{tabular}{rcrr}
            \hline
            Index & Term &   $\partial_t u_1$ &   $\partial_t u_2$ \\
            \hline
            1 & $u_{1}^{1}$                                                     &  1.0 &  0        \\
            2 & $u_{2}^{1}$                                                     &  0        &  1.0 \\
            6 & $u_{1}^{3}$                                                     & -1.0 &  4.0  \\
            7 & $u_{1}^{2}\,u_{1}^{1}$                                          & -4.0  & -1.0 \\
            8 & $u_{1}^{1}\,u_{1}^{2}$                                          & -1.0 &  4.0  \\
            9 & $u_{2}^{3}$                                                     &-4.0    & -1.0 \\
            28 & $\partial_x^{2}\left(u_{1}^{1}\right)$                          &  1.0 &  0        \\
            29 & $\partial_y^{2}\left(u_{1}^{1}\right)$                          &  1.0 &  0        \\
            31 & $\partial_x^{2}\left(u_{2}^{1}\right)$                          &  0        &  1.0 \\
            32 & $\partial_y^{2}\left(u_{2}^{1}\right)$                          &  0        &  1.0 \\
            \hline
        \end{tabular}
    \end{minipage}
    
    \vfill
    \begin{minipage}[t]{0.9\textwidth}
        \vspace{10pt}
        \hrule
        \vspace{5pt}
        45 Discarded Terms:\\
        $1.0$, $u_{1}^{2}$, $u_{1}^{1}\,u_{2}^{1}$, $u_{2}^{2}$, $\partial_x^{1}\left(u_{1}^{1}\right)$, $\partial_y^{1}\left(u_{1}^{1}\right)$, $\partial_x^{1}\left(u_{2}^{1}\right)$, $\partial_y^{1}\left(u_{2}^{1}\right)$, $\partial_x^{1}\left(u_{1}^{2}\right)$, $\partial_y^{1}\left(u_{1}^{2}\right)$, $\partial_x^{1}\left(u_{1}^{1}\,u_{2}^{1}\right)$, $\partial_y^{1}\left(u_{1}^{1}\,u_{2}^{1}\right)$, $\partial_x^{1}\left(u_{2}^{2}\right)$, $\partial_y^{1}\left(u_{2}^{2}\right)$, $\partial_x^{1}\left(u_{1}^{3}\right)$, $\partial_y^{1}\left(u_{1}^{3}\right)$, $\partial_x^{1}\left(u_{1}^{1}\,u_{2}^{2}\right)$, $\partial_y^{1}\left(u_{1}^{1}\,u_{2}^{2}\right)$, $\partial_x^{1}\left(u_{1}^{2}\,u_{2}^{1}\right)$, $\partial_y^{1}\left(u_{1}^{2}\,u_{2}^{1}\right)$, $\partial_x^{1}\left(u_{2}^{3}\right)$, $\partial_y^{1}\left(u_{2}^{3}\right)$, $\partial_x^{1}\partial_y^{1}\left(u_{1}^{1}\right)$, $\partial_x^{1}\partial_y^{1}\left(u_{2}^{1}\right)$, $\partial_x^{2}\left(u_{1}^{2}\right)$, $\partial_y^{2}\left(u_{1}^{2}\right)$, $\partial_x^{1}\partial_y^{1}\left(u_{1}^{2}\right)$, $\partial_x^{2}\left(u_{1}^{1}\,u_{2}^{1}\right)$, $\partial_y^{2}\left(u_{1}^{1}\,u_{2}^{1}\right)$, $\partial_x^{1}\partial_y^{1}\left(u_{1}^{1}\,u_{2}^{1}\right)$, $\partial_x^{2}\left(u_{2}^{2}\right)$, $\partial_y^{2}\left(u_{2}^{2}\right)$, $\partial_x^{1}\partial_y^{1}\left(u_{2}^{2}\right)$, $\partial_x^{2}\left(u_{1}^{3}\right)$, $\partial_y^{2}\left(u_{1}^{3}\right)$, $\partial_x^{1}\partial_y^{1}\left(u_{1}^{3}\right)$, $\partial_x^{2}\left(u_{1}^{1}\,u_{2}^{2}\right)$, $\partial_y^{2}\left(u_{1}^{1}\,u_{2}^{2}\right)$, $\partial_x^{1}\partial_y^{1}\left(u_{1}^{1}\,u_{2}^{2}\right)$, $\partial_x^{2}\left(u_{1}^{2}\,u_{2}^{1}\right)$, $\partial_y^{2}\left(u_{1}^{2}\,u_{2}^{1}\right)$, $\partial_x^{1}\partial_y^{1}\left(u_{1}^{2}\,u_{2}^{1}\right)$, $\partial_x^{2}\left(u_{2}^{3}\right)$, $\partial_y^{2}\left(u_{2}^{3}\right)$, $\partial_x^{1}\partial_y^{1}\left(u_{2}^{3}\right)$\\
        \vspace{5pt}
        \hrule
    \end{minipage}

    \caption{Full results for estimating reaction diffusion equations (\ref{eqn:rdc_1}) from data shown in figure~\ref{fig:rd_chaotic_2} from section~\ref{ssec:rd_chaotic} with additional thresholds for the terms $\partial_x^2u_1, \partial_y^2u_0$ and $\partial_x^2u_2, \partial_y^2u_1$. Found hyperparameters (thresholds): $h_1 = 0.3066755436690512$ for all terms for $\partial_t u_1$ except indices $28$ and $29$, $h_2=0.17758353931933746$ for all terms for $\partial_t u_2$ except indices $31$ and $32$, and ${h_{\text{sub}}}_1=0.005074209512766352$ for terms with indices $28$ and $29$ for $\partial_t u_1$, ${h_{\text{sub}}}_2= 0.0006699763535643887$ for terms with indices $31$ and $32$ for $\partial_t u_2$.}
    \label{tab:apx_rd_chaotic_2b}
\end{table}

\begin{table}[]
    \begin{minipage}[t]{0.48\textwidth}
        \centering
        Estimated model \\
        \vspace{5pt}
        \begin{tabular}{rcr}
        \hline
           Index & Term                                             &   $\partial_t u_1$ \\
        \hline
               1 & $u_{1}^{1}$                                      &  0.976494 \\
               7 & $\partial_x^{2}\left(u_{1}^{1}\right)$           &  1.01892  \\
               17 & $u_{1}^{1}\,{u_{\tau}}_{1}^{1}$                  & -0.97902  \\
        \hline
        \end{tabular}
    \end{minipage}
    \hfill
    \begin{minipage}[t]{0.48\textwidth}
        \centering
        True model ($\tau = 1.0$) \\
        \vspace{5pt}
        \begin{tabular}{rcr}
        \hline
           Index & Term                                             &   $\partial_t u_1$ \\
        \hline
               1 & $u_{1}^{1}$                                      &  1.0 \\
               7 & $\partial_x^{2}\left(u_{1}^{1}\right)$           &  1.0  \\
               17 & $u_{1}^{1}\,{u_{\tau}}_{1}^{1}$                  & -1.0  \\
        \hline
        \end{tabular}
    \end{minipage}
    \vfill
    \begin{minipage}[t]{0.9\textwidth}
        \vspace{10pt}
        \hrule
        \vspace{5pt}
        20 Discarded Terms:\\
        $1.0$, $u_{1}^{2}$, $u_{1}^{3}$, $\partial_x^{1}\left(u_{1}^{1}\right)$, $\partial_x^{1}\left(u_{1}^{2}\right)$, $\partial_x^{1}\left(u_{1}^{3}\right)$, $\partial_x^{2}\left(u_{1}^{2}\right)$, $\partial_x^{2}\left(u_{1}^{3}\right)$, $\partial_x^{3}\left(u_{1}^{1}\right)$, $\partial_x^{3}\left(u_{1}^{2}\right)$, $\partial_x^{3}\left(u_{1}^{3}\right)$, $\partial_x^{4}\left(u_{1}^{1}\right)$, $\partial_x^{4}\left(u_{1}^{2}\right)$, $\partial_x^{4}\left(u_{1}^{3}\right)$, ${u_{\tau}}_{1}^{1}$, $u_{1}^{2}\,{u_{\tau}}_{1}^{1}$, $\partial_x^{ 1}\left({u_{\tau}}_{1}^{1}\right)$, $\partial_x^{ 2}\left({u_{\tau}}_{1}^{1}\right)$, $\partial_x^{ 3}\left({u_{\tau}}_{1}^{1}\right)$, $\partial_x^{ 4}\left({u_{\tau}}_{1}^{1}\right)$\\
        \vspace{5pt}
        \hrule
        
    \end{minipage}

    \caption{Full results for estimating the time-delayed Fisher-KPP equation~(\ref{eqn:fisher_kpp}) from data shown in figure~\ref{fig:fisherKPP} from section~\ref{ssec:fisherkpp}. Found hyperparameter (threshold and time-delay): $h_1 = 0.6238713362317317$ and $\tau = 1.0172462526772437$.}
    \label{tab:apx_fkpp}
\end{table}

\end{document}